\documentclass[twoside,twocolumn]{revtex4-1}
\usepackage{lmodern}
\usepackage{lmodern}
\usepackage[T1]{fontenc}
\usepackage[utf8]{inputenc}
\usepackage{color}
\usepackage{array}
\usepackage{textcomp}
\usepackage{mathrsfs}
\usepackage{amsmath}
\usepackage{amssymb}
\usepackage{graphicx}
\PassOptionsToPackage{version=3}{mhchem}
\usepackage{mhchem}
\usepackage[unicode=true,
 bookmarks=false,
 breaklinks=false,pdfborder={0 0 1},backref=false,colorlinks=false]
 {hyperref}
\hypersetup{
 hidelinks}

\makeatletter

\providecommand{\tabularnewline}{\\}

\usepackage{placeins}
\usepackage{units}
\usepackage[T1]{fontenc} 
\usepackage{amsmath}
\usepackage{mathtools}
\allowdisplaybreaks 
\usepackage{amssymb}
\usepackage{graphicx}
\usepackage[bf, small, centerlast, font=footnotesize]{caption}
\usepackage{wrapfig}
\usepackage{footnote}
\usepackage{listings}
\usepackage{color}
\usepackage[wide]{sidecap}
\usepackage{epstopdf}
\usepackage{marvosym}
\usepackage{hyperref}
\usepackage{bbm}

\renewcommand{\epsilon}{\varepsilon}
\renewcommand{\phi}{\varphi}
\renewcommand{\theta}{\vartheta}
\renewcommand{\rho}{\varrho}

\makeatother

\begin{document}

\title{MSM/RD: Coupling Markov state models of molecular kinetics with reaction-diffusion
simulations}

\author{Manuel Dibak$^{1,\dagger}$, Mauricio J. del Razo$^{1,\dagger}$,
David De Sancho$^{2}$, Christof Schütte$^{1}$ and Frank Noé$^{1,a)}$}
\begin{abstract}
\textbf{Abstract:} \textcolor{black}{Molecular dynamics (MD) simulations
can model the interactions between macromolecules with high spatiotemporal
resolution but at a high computational cost. By combining high-throughput
MD with Markov state models (MSMs), it is now possible to obtain long-timescale
behavior of small to intermediate biomolecules and complexes. To model
the interactions of many molecules at large lengthscales, particle-based
reaction-diffusion (RD) simulations are more suitable but lack molecular
detail. Thus, coupling MSMs and RD simulations (MSM/RD) would be highly
desirable, as they could efficiently produce simulations at large
time- and lengthscales, while still conserving the characteristic
features of the interactions observed at atomic detail. While such
a coupling seems straightforward, fundamental questions are still
open: Which definition of MSM states is suitable? Which protocol to
merge and split RD particles in an association/dissociation reaction
will conserve the correct bimolecular kinetics and thermodynamics?
In this paper, we make the first step towards MSM/RD by laying out
a general theory of coupling and proposing a first implementation
for association/dissociation of a protein with a small ligand ($A+B\rightleftharpoons C$).
Applications on a toy model and CO diffusion into the heme cavity
of myoglobin are reported.}
\end{abstract}
\maketitle
\noindent \vspace{-0.5cm}

\noindent $^{\dagger}$Equal contribution

\noindent $^{1}$Freie Universität Berlin, Department of Mathematics
and Computer Science, Arnimallee 6, 14195 Berlin, Germany

\noindent $^{2}$Kimika Fakultatea, Euskal Herriko Unibertsitatea
(UPV/EHU), and Donostia International Physics Center (DIPC), P.K.
1072, 20080 Donostia, Euskadi, Spain

\noindent $^{a)}$Corresponding author. Electronic mail: frank.noe@fu-berlin.de.

\section{Introduction}

Life processes such as cellular signaling, control and regulation
arise from complex interactions and reactions between biomolecules.
A fundamental challenge of understanding and controlling life processes
is that they are inherently multiscale – cellular signaling alone
involves $6$ orders of magnitude in lengthscales ($0.1$ nanometers
to $100$ micrometers) and $18$ orders of magnitude in timescales
(femtoseconds to hours). Unfortunately, these scales are tightly coupled
– a single-point mutation in a protein can disturb the biochemical
interactions such that this results in disease or death of the organism.
No single experimental or simulation technique can probe all time-
and lengthscales at a resolution required to understand such a process
comprehensively. 

In computer simulations, this dilemma can be mitigated by multiscale
techniques – different parts of the system are described by a high-resolution
and a low-resolution model, and these parts are coupled to give rise
to a hybrid simula\textcolor{black}{tion. A famous example of such
a multiscale model in biophysical chemistry is the coupling of quantum
mechanics and molecular mechanics (QM/MM) \citep{WarshelLevitt_JMB76_QMMM}.
Here we lay the foundations for a hybrid simulation technique that
couples two scales that are particularly useful to model intracellular
dynamics: a Markov state model (MSM) of the molecular dynamics (MD)
scale that describes structural changes of biomolecules and their
complexes, and the reaction-diffusion scale that describes diffusion,
association and dissociation on the lengthscale of a cell. We call
this approach MSM/RD, due to the combination of the simulation models
chosen at these scales:}
\begin{enumerate}
\item \textcolor{black}{MSMs of the molecular scale: MD simulation allows
us to probe molecular processes at atomic detail, but its usefulness
has long been limited by the sampling problem. Recently, the combination
of hard- and software for high-throughput MD simulations \citep{ShirtsPande_Science2000_FoldingAtHome,BuchEtAl_JCIM10_GPUgrid,Shaw_Science10_Anton,DoerrEtAl_JCTC16_HTMD}
with MSMs \citep{PrinzEtAl_JCP10_MSM1,BowmanPandeNoe_MSMBook,SarichSchuette_MSMBook13}
has enabled the extensive statistical description of p}rotein folding
and conformation changes \citep{NoeSchuetteReichWeikl_PNAS09_TPT,Bowman_JCP09_Villin,LindorffLarsenEtAl_Science11_AntonFolding,KohlhoffEtAl_NatChem14_GPCR-MSM},
as well as the association of proteins with ligands \citep{BuchFabritiis_PNAS11_Binding,SilvaHuang_PlosCB_LaoBinding,PlattnerNoe_NatComm15_TrypsinPlasticity,deSancho2015identification,kubas2016mechanism}
and even other proteins \citep{PlattnerEtAl_NatChem17_BarBar}. Using
multi-ensemble Markov models (MEMMs) \citep{WuMeyRostaNoe_JCP14_dTRAM,RostaHummer_DHAM,WuEtAL_PNAS16_TRAM,MeyWuNoe_xTRAM},
MSMs can be derived that even capture the kinetics of ultra-rare events
beyond the seconds timescale at atomistic resolution \citep{PaulEtAl_PNAS17_Mdm2PMI,CasasnovasEtAl_JACS17_UnbindingKinetics}.
MSM approaches can thus model the long-lived states and transition
rates of molecular detail interactions, but the cost of atomistic
MD sampling limits them to relatively small biomolecules and complexes. 
\item Reaction-diffusion (RD) scale: While atomic detail is relevant for
some processes that affect the cellular scale, it is neither efficient
nor insightful to maintain atomic resolution at all times for cellular
processes. We choose particle-based reaction-diffusion (PBRD) dynamics
kinetics as a reference model for the cellular scale. PBRD simulates
particles, representing individual copies of proteins, ligands or
other metabolites. Particles move in space via diffusion and reactive
species will react with a probability according to their reaction
rate when being clo\textcolor{black}{se by. Here, a reaction may represent
molecular processes such as binding, dissociation, conformational
change, or actual enzymatic reactions. PBRD acknowledges that chemical
reactions are inherently discrete and stochastic in nature \citep{qian2010cellular},
and that diffusion in cells is often not fast enough to justify well-stirred
reaction kinetics \citep{erban2009stochastic,fange2010stochastic,takahashi2010spatio}.
A large number of recent software packages and codes implement some
form of PBRD \citep{andrews2004stochastic,BiedermannEtAl_BJ15_ReaddyMM,donev2010first,donev2017efficient,hattne2005stochastic,SchoenebergNoe_PlosOne13_ReaDDy,van2005green,ZonTenWolde_PRL05_GFRD},
see also the reviews \citep{mereghetti2011diffusion,SchoenebergUllrichNoe_BMC14_RDReview}.
Hydrodynamic interactions at this scale could be incorporated by particle-based
coupling terms \citep{ermak1978brownian,geyer2009n}. The effect of
crowders and complicated boundaries such as membranes on the particle
diffusion can be represented by including interaction forces on the
RD scale \citep{SchoenebergNoe_PlosOne13_ReaDDy}.}
\end{enumerate}
\textcolor{black}{In the limit that the conformational transitions
of all molecules are fast, the MSM dynamics of each molecule effectively
averages, and the interaction between the molecules (e.g. association)
occurs with suitably averaged rates, reducing the problem to PBRD.
However, when the lifetimes of some conformations are long compared
to the typical time between two molecular interactions, or even the
time between successive rebinding events of two molecules, the conformation
dynamics of molecules described by the MSM part couples with the RD
dynamics. MSM/RD opens up the possibility to simulate and analyze
such effects quantitatively. For example, bimolecular binding rates
from MD-derived MSMs can be inaccurate due to periodic boundary effects
and a short-lived dissociated state in comparison to the MSM lag-time
\citep{PlattnerEtAl_NatChem17_BarBar}. MSM/RD can overcome these
issues by extending the diffusion domain available lessening the periodic
boundary effects and increasing the lifetime of the dissociated state. }

\textcolor{black}{The ultimate aim of MSM/RD is to produce an efficient
multiscale simulation that reproduces the essential statistical behavior
of a practically unaffordable large-scale MD simulation by employing
only statistics obtained from simulations of the constituent biomolecules
in small solvent boxes. As developing a full theory involving rotational
diffusion, three- or more-body interactions, hydrodynamics will be
highly complex, we here aim to make a first step towards this goal
by coupling MSM and RD scales for bimolecular systems without large-scale
hydrodynamic interactions.}

\begin{figure}
\centering 

\includegraphics[width=1\columnwidth]{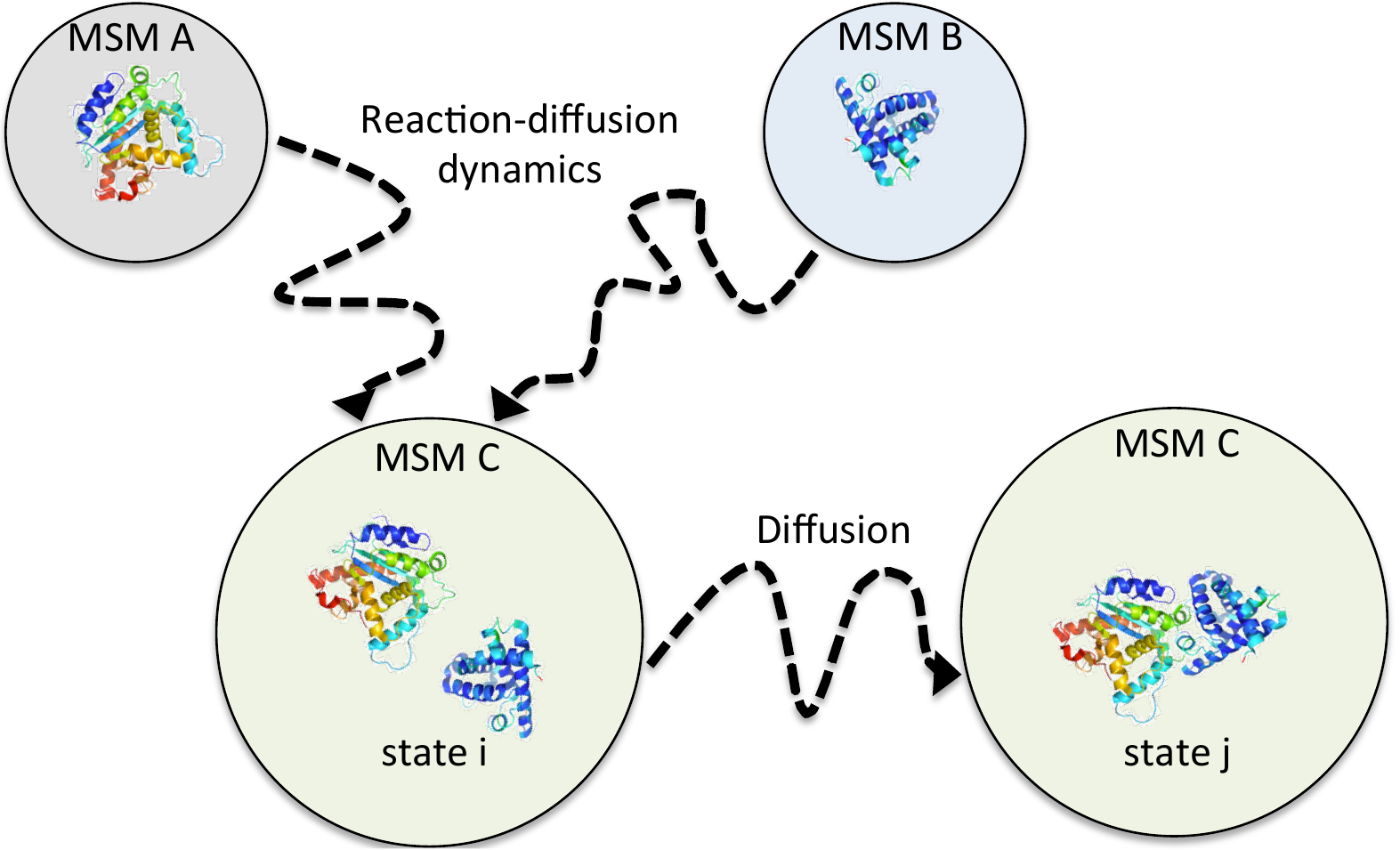}

\caption{Sketch of the MSM/RD scheme. When molecules $A$ and\textcolor{black}{{}
$B$ are not in close proximity, they diffuse freely. When $A$ and
$B$ are close, they merge into a complex particle $C$ which itself
diffuses and whose internal dynamics are encoded by coupled MSM state
transitions. When the molecules transition into a dissociated state,
they are again separated into two separately diffusing particles $A$
and $B$ with initial positions depending on the last MSM state. Note
that in the dissociated state, molecules $A$ and $B$ could also
potentially undergo conformational changes encoded in independent
MSM state transitions.}}

\label{fig:schemeMSMRD-myoglobin} 
\end{figure}

\textcolor{black}{We first derive a theory of MSM/RD for bimolecular
systems, as depicted in Fig. \ref{fig:schemeMSMRD-myoglobin}. When
the two molecules are far from each other, they both undergo a diffusion
process. When they come close to each other, molecular interactions,
modeled with MD-derived MSMs, need to be taken into account. We further
develop an algorithm to couple the MSM and RD scales for the special
case of a protein interacting with a ligand, which is one of the main
advances in this paper. This is not a trivial undertaking since one
needs to solve two problems: to couple the MSM and RD part in such
a way that the correct macroscopic rates and equilibrium probabilities
are recovered, and to develop a suitable MSM discretization such that
this coupling can be made. We demonstrate the validity of our theory
and algorithms on a toy model of p}rotein-ligand interaction and on
binding of carbon monoxide to myoglobin.

In related work, \citep{sbailo2017efficient,vijaykumar2015combining}
have coupled MD with a diffusion scheme. The work \citep{VotapkaAmaro_SEEKR_SPCB17}
further incorporates milestoning theory \citep{faradjian2004computing}
to compute the local kinetic information in terms of transitions between
milestones via short MD runs. In contrast with their work, we do not
employ direct MD simulations at the ``small'' scale, but represent
the small scale by an MSM as this allows us to operate on roughly
the same timesteps for the small and the large scales. Other works
have proposed alternative schemes to couple random walks (MSMs) with
Brownian diffusion schemes, some examples can be seen in \citep{del2016discrete,flegg2012two,flegg2015convergence}.
However, these works focus on specific contexts that are not directly
applicable for coupling MD-derived MSMs with reaction-diffusion schemes.

\section{MSM/RD: coupling Markov state models and reaction-diffusion}

\textcolor{black}{We develop a theoretical description for MSM/RD.
The relevant scenarios for MSM/RD can be classified by the number
of interacting particles, or the related reaction order:}
\begin{enumerate}
\item \textcolor{black}{First-order reactions: isolated diffusing particles
can be modeled by an MSM obtained from MD simulations in a solvent
box. The MSM directly translates into a set of unimolecular reactions
that can be implemented in standard PBRD software. As long as the
particles don't interact, the only effect of different states on the
dynamics are changes between different diffusion constants/tensors.}
\item \textcolor{black}{Second-order reactions: interactions between two
molecules that can be modeled as bimolecular reactions including protein-ligand
or protein-protein association ($A+B\rightarrow C$). As soon as the
complex $C$ has been formed, its dynamics may be described by state
transitions of an MSM of the complex.}
\item \textcolor{black}{Higher-order reactions: simultaneous interactions
between more than two molecules. }
\end{enumerate}
\textcolor{black}{In this work, we will focus on second-order reactions.
First-order reactions are trivial state changes of a particle that
are occurring as part of the MSM dynamics. Consistent with current
conventions in PBRD frameworks, we follow the convention of breaking
down higher-order reactions to second-order reactions, although in
Sec. \ref{sec:Conclusion} we suggest possible extensions to treat
these explicitly.}

\textcolor{black}{In order to derive the theory for second-order reactions,
we concentrate on the dynamics of two molecules, $A$ and $B$. For
the sake of simplicity, we assume the two molecules do not have conformational
changes of their own, so they can only diffuse and interact with each
other. However, it is straightforward to extend MSM/RD to include
conformational changes (first-order reactions) coupled with second-order
reactions.}

\subsection{The \emph{ground truth} model with full dynamics}

\textcolor{black}{\emph{Ground truth}}\textcolor{black}{{} is a term
often used in machine learning that refers to a reference model with
respect to which modeling errors are measured. In the present context,
the }\textcolor{black}{\emph{ground truth}}\textcolor{black}{{} model
contains the two (or more) solute molecules whose interactions will
be later approximated by an MSM in a }\textcolor{black}{\emph{large-scale}}\textcolor{black}{{}
simulation, i.e. a simulation box that is not truncated after a small
solvent boundary as customary for MD simulation. Importantly, there
is no universally correct ground truth, but this model employs the
MD simulation setup and dynamical model chosen by the user for the
modeling task at hand. This choice includes the MD force field, solvation
conditions and ion concentration, the protonation state at the pH
of interest or even constant-pH simulations \citep{DonniniEtAl_JCTC2011_ConstantPH},
the treatment of electrostatics, the thermostat, the integrator and
time step, etc.}

\textcolor{black}{If such a large-scale model were simulated for a
long time or with many trajectories, it would give rise to statistical
properties of the solute molecules that we want to reproduce, such
as their equilibrium constants and association rates. However, such
a simulation is in general inefficient or infeasible, and our aim
is that to reproduce its statistical properties using an MSM/RD model
that is parametrized only using small MD simulations of the constituent
solute molecules and complexes.}

\textcolor{black}{For simplicity, we derive the MSM/RD theory using
all-atom explicit solvent MD simulations with a Langevin thermostat
as the ground truth, as this setup is frequently used for MD simulations.
However, the MSM/RD results apply more generally, e.g. to different
choices of thermostats or integrators, as the MSM limit for long-time
description of the dynamics and the overdamped limit for long-time
and large-scale description of the solute transport are achieved from
a large family of ground truth models.}

\textcolor{black}{Langevin dynamics evolve as:}
\begin{equation}
m_{k}\frac{d^{2}}{dt^{2}}x_{k}(t)=-\nabla_{k}U(\mathbf{x}_{t})-\gamma_{k}\frac{d}{dt}x_{k}+\sqrt{2k_{B}T\gamma_{k}}\boldsymbol{\xi}_{k}(t),\label{eq:Langevin}
\end{equation}
where $x_{k}$ represents the three-dimensional position of the $k^{th}$
atom in the system (including the solvent), $\mathbf{x}_{t}=[x_{1}(t),\dots,x_{k}(t),\dots,x_{N}(t)]$,
$N$ the total number of atoms, $U$ is the potential energy and $-\nabla_{k}U$
is the force acting on the $k^{th}$ particle, $m_{k}$ is the $k^{th}$
particle mass, $\gamma_{k}$ is the $k^{th}$ damping coefficient,
and $\boldsymbol{\xi}_{k}(t)$ is a Gaussian random force such that
the expectations of its components satisfy $E[\xi_{k,i}(t)]=0$ (zero
mean) and $E[\xi_{k,i}(t)\xi_{k,j}(s)]=\delta_{ij}\delta(t-s)$ (white
noise) with $k_{B}T$ being the thermal energy. In simulations, we
use finite-time-step approximations of (\ref{eq:Langevin}) and use
it to generate stochastic trajectories. For the theoretical analysis,
it is more useful to look at the ensemble dynamics, i.e., the propagation
of probability densities in time. For this, we can ask: If we start
the dynamical system in phase space point $\mathbf{y}$ and let it
run, with which probability will we find it in a point $\mathbf{x}$
a time $\tau$ later? We call this probability the transfer probability
$p(\mathbf{y}\rightarrow\mathbf{x};\,\tau)$, and we will use it to
describe the action of the ground truth dynamics \citep{SchuetteFischerHuisingaDeuflhard_JCompPhys151_146}.
The transfer probability $p(\mathbf{y}\rightarrow\mathbf{x};\,\tau)$
subsumes the full complexity of the MD model, in\textcolor{black}{cluding
interaction energies of the molecules with each other and external
fields, and it can be constructed regardless of which thermostat or
integrator is used. The propagation of prob}ability densities $\rho(\mathbf{x};\,t)$
in time is formally described by the propagator $\mathcal{P}_{\tau}$:
\begin{align}
\rho(\mathbf{x};\,t+\tau) & =\mathcal{P}_{\tau}\rho(\mathbf{x};\,t)\nonumber \\
 & =\int p(\mathbf{y}\rightarrow\mathbf{x};\,\tau)\rho(\mathbf{y};\,t)\,\mathrm{d}\mathbf{y}\label{eq:propagation_propagator}
\end{align}
We want to find an efficient algorithm to approximate these dynamics.
More specifically we want to approximate certain aspects of these
dynamics, such as the long-time behavior.

It is often useful to consider densities relative to the stationary
density $\pi(\mathbf{x})$ given by
\[
u(\mathbf{x};\,t)=\frac{\rho(\mathbf{x};\,t)}{\pi(\mathbf{x})},
\]
which defines the propagator relative to the stationary density, or
transfer operator \citep{SchuetteFischerHuisingaDeuflhard_JCompPhys151_146}:
\begin{align}
u(\mathbf{x};\,t+\tau) & =\mathcal{T}_{\tau}u(\mathbf{x};\,t)\nonumber \\
 & =\int\frac{\pi(\mathbf{y})}{\pi(\mathbf{x})}p(\mathbf{y}\rightarrow\mathbf{x};\,\tau)u(\mathbf{y};\,t)\,\mathrm{d}\mathbf{y}\nonumber \\
 & =\int p(\mathbf{x}\rightarrow\mathbf{y};\,\tau)u(\mathbf{y};\,t)\,\mathrm{d}\mathbf{y}\label{eq:propagation_transfer_operator}
\end{align}
The third row follows from detailed balance. For reversible systems,
where detailed balance is fulfilled, $\mathcal{T}_{\tau}$ is often
called backward propagator, as it appears to evolve densities backward
in time.

We will now introduce a scale separation by treating molecules $A$
and $B$ different when they are close (interacting) and far apart
(non-interacting). More specifically these scales are defined by the
distance between the centers of mass of $A$ and $B$, $r_{AB}$:
\begin{enumerate}
\item MSM domain: molecules are in the \emph{interaction} region $I=\{\mathbf{x}\mid r_{AB}(\mathbf{x})<R\}$. 
\item RD domain: molecules are in the \emph{outside} region $O=\{\mathbf{x}\mid r_{AB}(\mathbf{x})\ge R\}$.
\end{enumerate}
The definition of $R$ will be investigated later. Next, we take a
closer look at the dynamics valid in these respective domains.

\subsection{Markov state models for interacting molecules\label{subsec:innerMSM}}

We consider molecules that are closer than $R$ to be interacting,
hence we call the corresponding subset of state space $I$. The kinetics
in $I$ are fully described by Eq. (\ref{eq:propagation_transfer_operator}),
which can be approximat\textcolor{black}{ed by an MSM derived from
a MD simulation that fully includes $I$ (usually plus some extra
space, because MD simulations typically employ periodic rather than
spherical boundary conditions). We implicitly assume that the interaction
forces between  proteins or protein-ligand pairs have decayed to zero
at distances $R$ or greater. Note that this assumption requires that
the MD simulation conducted to parametrize an MSM/RD model has a sufficiently
large simulation box, suitable electrostatics treatment and solvation
conditions (ions etc.) such that in the dissociated state the solutes
can be in any orientation without significantly interacting with each
other or with their periodic images.}

\textcolor{black}{The interaction region $I$ will here be approximated
by an MSM. We perform a spectral decomposition of (\ref{eq:propagation_transfer_operator}),
assuming that there exists a unique stationary density $\pi$ and
the dynamics obey detailed balance. Furthermore, we truncate the spectral
decomposition after a finite number of $k$ terms:
\begin{equation}
\rho(\mathbf{x};\,t+\tau)\approx\pi(\mathbf{x})\sum_{i=1}^{k}\lambda_{i}(\tau)\langle\psi_{i}(\mathbf{x}),\rho(\mathbf{x};\,t)\rangle\psi_{i}(\mathbf{x})\label{eq:transfSpectral}
\end{equation}
Here, $\langle,\rangle$ denotes the scalar product with respect to
$\pi(\mathbf{x})$, $\psi_{i}$ are the eigenfunctions of $\mathcal{T}_{\tau}$
and its leading eigenvalues have the form
\[
\lambda_{i}(\tau)=\mathrm{e}^{-\tau/t_{i}},
\]
where $t_{i}$ is a characteristic relaxation timescale. The truncation
after $k$ terms in Eq. (\ref{eq:transfSpectral}) assumes that $\tau$
is long compared to $t_{k+1}$, where $3t_{k}<\tau$ is sufficient
for practical purposes. Most of the solvent dynamics correspond to
the fast coordinates that are averaged out (large-scale hydrodynamics
are not part of the MSM term, while solvent molecules with long-lived
interactions with the solute molecule can be considered to be part
of that solute molecule). Now we can perform a Galerkin projection
of the transfer operator by discretizing the phase space using basis
functions $\chi_{i}(\mathbf{x}),\,i=1,...,n$. In MSMs, these are
characteristic functions 
\[
\chi_{i}(\mathbf{x})=\begin{cases}
1 & \mathbf{x}\in S_{i}\\
0 & \mathbf{x}\notin S_{i}
\end{cases}
\]
where the $S_{i}$ form a complete partition of phase space, i.e.
$\Omega=\{S_{1}\cup S_{2}\cup\cdots\cup S_{n}\}$. The phase space
has now been discretized into a finite state space. The local densities
become vectors simply given by
\[
\pi_{j}=\int_{x\in S_{j}}\pi(\mathbf{x})d\mathbf{x},\qquad\rho_{j}(t)=\int_{x\in S_{j}}\rho(\mathbf{x};\,t)d\mathbf{x}.
\]
Furthermore, we want the transfer operator to be approximated by a
matrix. We can obtain this matrix by noting that the eigenfunctions
of the transfer operator also become vectors in state space}

\textcolor{black}{
\[
\psi_{i}^{j}=\frac{1}{\int_{\Omega}\chi_{j}(\mathbf{x})d\mathbf{x}}\int_{x\in S_{j}}\psi_{i}(\mathbf{x})d\mathbf{x}.
\]
Inserting these equations into (\ref{eq:transfSpectral}), and rewriting
it in matrix form, we obtain the Chapman-Kolmogorow equation}

\textcolor{black}{
\begin{equation}
\boldsymbol{\rho}(t+\tau)=\mathbf{T}^{\top}(\tau)\,\boldsymbol{\rho}(t),\label{eq:MSM}
\end{equation}
with $\lambda_{i}(\tau)$ and $\boldsymbol{\psi_{i}}=[\psi_{i}^{1},\dots,\psi_{i}^{n}]$
the $i^{th}$ eigenvalue and eigenvector of the transition probability
matrix $\mathbf{T}(\tau),$ respectively, and with $\boldsymbol{\rho}(t)=[\rho_{1}(t),\dots,\rho_{n}(t)]$
the probability mass function \citep{PrinzEtAl_JCP10_MSM1}.}

\textcolor{black}{Estimating a high-quality MSM from MD simulation
data can be quite complex. It typically involves (i) mapping the MD
coordinates to a set of features, such as residue distances, contact
maps or torsion angles, (ii) reducing the dimension to slow collecti}ve
variables (CVs), often based on the variational approach or conformation
dynamics \citep{NoeNueske_MMS13_VariationalApproach,NueskeEtAl_JCTC14_Variational}
or its special case time-lagged independent component analysis (TICA)
\citep{PerezEtAl_JCP13_TICA,SchwantesPande_JCTC13_TICA} – see \citep{NoeClementi_COSB17_SlowCVs,KlusEtAl_JNS17_DataDriven}
for an overview, (iii) optionally, embedding the resulting coordinates
in a metric space whose distances correspond to some form of dynamical
distance \citep{NoeClementi_JCTC15_KineticMap,NoeClementi_JCTC16_KineticMap2},
(iv) discretizing the result space using data-based clustering \citep{BowmanPandeNoe_MSMBook,SchererEtAl_JCTC15_EMMA2,HarriganEtAl_BJ17_MSMbuilder},
typically resulting in 100-1000 discrete states, and (v) estimating
the transition matrix $\mathbf{T(\tau)}$ or a transition rate matrix
$\mathbf{K}$ with $\mathbf{T}(\tau)=\exp\left(\tau\mathbf{K}\right)$
at some lag time $\tau$, and validating it \citep{Bowman_JCP09_Villin,PrinzEtAl_JCP10_MSM1,BucheteHummer_JPCB08,TrendelkampSchroerEtAl_InPrep_revMSM}.
Finally, the MSM may be coarse-grained to few metastable states \citep{KubeWeber_JCP07_CoarseGraining,HummerSzabo_JPCB15_CoarseGraining,OrioliFaccioli_JCP16_CoarseMSM,NoeEtAl_PMMHMM_JCP13}.
The MSM software packages PyEMMA \citep{SchererEtAl_JCTC15_EMMA2}
and MSMbuilder \citep{HarriganEtAl_BJ17_MSMbuilder} can greatly help
to simplify this process and make it reproducible.

In the case where there are well-defined meta-stable regions in phase\textcolor{black}{{}
space, we can greatly reduce the number of states in the MSM. One
way to simplify the MSM construction process above and to directly
end up with a few-state MSM is to employ VAMPnets, where the complex
MSM construction pipeline is replaced by a neural network that is
trained using the variational approach for Markov processes \citep{MardtEtAl_VAMPnets}.
Alternatively, one can replace the discretization step (iv) above
by employing a core set approach that was derived in \citep{BucheteHummer_JPCB08}
and further analyzed in \citep{schutte2011markov}. The essential
idea is to define the states as cores around the metastable regions.
Due to the metastability, the probability of finding the system outside
of the metastable regions is very small, so to a good approximation
the kinetics can be described as a core-to-core jump process \citep{schutte2011markov}.
This approach will be employed throughout this paper and and explained
in more detail in Sec. \ref{sec:MSM/RD-implementation}.}

\subsection{Reaction-diffusion dynamics for noninteracting molecules}

When molecules are far apart, and thus in the RD domain defined by
$r_{AB}(\mathbf{x})\ge R$, they are not directly interacting. As
the dynamics of the two molecules are independent, it is convenient
to only track the net diffusion of the centers of mass, $\boldsymbol{r}_{A}$
and $\boldsymbol{r}_{B}$. Furthermore, we assume that the dynamics
in the RD domain can be tracked by coarse timesteps of at least $\Delta t$
which exceeds the typical velocity autocorrelation time (picoseconds).
At such timescales, the fast dynamics corresponding to the solvent
are averaged out. It is possible that even longer timesteps are made
using an event-based integration scheme such as first-passage kinetic
Monte Carlo (FPKMC) algorithm, Green's function reaction dynamics
(GFRD) or MD-GFRD \citep{donev2010first,takahashi2010spatio,vijaykumar2015combining,van2005green,ZonTenWolde_PRL05_GFRD}.
At such timesteps, the Langevin equation (\ref{eq:Langevin}) becomes
an overdamped Langevin equation for the centers of mass of the two
molecules, i.e. the motion is governed by pure diffusion: 

\begin{equation}
\frac{d\boldsymbol{r}_{A}(t)}{dt}=\sqrt{2D_{A}}\boldsymbol{\xi}_{A}(t),\quad\frac{d\boldsymbol{r}_{B}(t)}{dt}=\sqrt{2D_{B}}\boldsymbol{\xi}_{B}(t),\label{eq:overdampMassCenters}
\end{equation}
where $\boldsymbol{\xi}_{A}(t)$ and $\boldsymbol{\xi}_{B}(t)$ are
independent white noise vectors with each of their components satisfying
$E[\xi_{K,i}(t)]=0$ and $E[\xi_{K,i}(t)\xi_{K,j}(s)]=\delta_{ij}\delta(t-s)$.
$D_{A}$ and $D_{B}$ are the net diffusion coefficients for the centers
of mass, which can be obtained from MD simulations. In general, as
we are tracking the center of mass, we also need to track the rotational
diffusion of the molecules. However, as rotational diffusion is not
relevant for the examples discussed in this manuscript, we refer to
\citep{SchluttigEtAl_JCP08_PatchyParticleAssociation,VijaykumarEtAl_Arxiv16_AnisotropicMultiscaleGFRD}. 

In the present case, we can simply fix the frame of reference in $\boldsymbol{r}_{A}(t)$,
assume the rotation of $A$ is slower than the diffusion of $B$,
which is true for protein-ligand systems, and fix the orientation
of the axis to that of molecule $A$. We further assume that $B$
is a small molecule such that its orientation is not very relevant,
as it will be the case in our implementation of the scheme. This simplifies
Eqs. \ref{eq:overdampMassCenters} into a simple diffusion in $\boldsymbol{r}_{B}$
only 

\begin{equation}
\frac{d\boldsymbol{r}_{B}(t)}{dt}=\sqrt{2(D_{B}+D_{A})}\boldsymbol{\xi}(t),\label{eq:COM_diffusion}
\end{equation}
with the components of $\boldsymbol{\xi}(t)$ satisfying $E[\xi_{i}(t)]=0$
and $E[\xi_{i}(t)\xi_{,j}(s)]=\delta_{ij}\delta(t-s)$. 

\subsection{MSM/RD coupled dynamics\label{subsec:MSM/RD-coupled}}

The present coupled model only considers interactions between up to
two molecules. This is a frequent assumption in PBRD \citep{ZonTenWolde_PRL05_GFRD,donev2010first,SchoenebergNoe_PlosOne13_ReaDDy,BiedermannEtAl_BJ15_ReaddyMM}
but may be restrictive from a molecular standpoint. We assume that
simultaneous reactions between three or more molecules such as $A+B+C\rightarrow D$
can always be broken down into $A+B\rightarrow AB;\,AB+C\rightarrow D$
or other bimolecular pathways, and therefore focus on MSM/RD involving
two molecules. In order to do the coupling, as the dynamics in the
$I$ and $O$ region are given in terms of states and coordinates
respectively, we need to recognize that $\mathbf{x}$ and $\mathbf{y}$
in the transfer density $p(\mathbf{y}\rightarrow\mathbf{x};\,\tau)$
can be either coordinates $\mathbf{c}$ (center-of-mass position and
perhaps orientation of the molecule) or states $s$ (metastable regions
in the coordinate space). In order to implement the coupling, we suggest
defining two quantities: 
\begin{itemize}
\item \texttt{$p_{\mathrm{entry}}\left[\mathbf{c}_{t}\rightarrow\mathbf{x}_{t+\Delta};\,\Delta\right]$},
transfer probability of starting in coordinates $\mathbf{c}_{t}$
just inside the MSM domain ($r_{AB}(\mathbf{c}_{t})<R$) conditioned
on hitting only one state $\mathbf{x}_{t+\Delta}=s_{t+\Delta}$ in
the MSM domain (transition event) OR on exiting once the MSM domain
$\mathbf{x}_{t+\Delta}=\mathbf{c}_{t+\Delta}$ (return event)
\item $p_{\mathrm{exit}}\left[s_{t}\rightarrow\mathbf{x}_{\mathrm{t}+\Delta};\,\Delta\right]$,
transfer probability of starting in state $s_{t}$ conditioned on
exiting once the MSM domain $\mathbf{x}_{t+\Delta}=\mathbf{c}_{t+\Delta}$
(exit event) OR hitting once any other state $(\mathbf{x}_{t+\Delta}=s_{t+\Delta})$. 
\end{itemize}
Once we know these transfer probabilities, we can introduce the basic
algorithm where $\tau_{\mathrm{RD}}$ and $\tau_{\mathrm{MSM}}$ correspond
to the diffusion and MSM time-step, respectively: 

\texttt{\noindent}\texttt{\textbf{Input}}\texttt{: Initial mode (RD
or MSM), initial condition (coordinates $\mathbf{c}_{0}$ or state
$s_{0}$, respectively) and $t=0$:}

\texttt{While $t\leq t_{\mathrm{final}}:$}
\begin{enumerate}
\item \texttt{If in RD mode:}
\begin{enumerate}
\item \texttt{Propagate $\mathbf{c}_{t}\rightarrow\mathbf{c}_{t+\tau_{\mathrm{RD}}}$
by diffusion}
\item \texttt{Update time $t\mathrel{{+}{=}}\tau_{\mathrm{RD}}$}
\item \texttt{If $r_{AB}(\mathbf{c}_{t})<R$ (enter MSM domain):}
\begin{itemize}
\item \texttt{Sample next event ($\mathbf{x}_{t+\Delta}$,$\Delta$)from}~\\
\texttt{ $p_{\mathrm{entry}}[\mathbf{c}_{t}\rightarrow\mathbf{x}_{t+\Delta};\,\Delta]$.}
\item \texttt{If transition event:}~\\
\texttt{Map to state }$s_{t+\Delta}=\mathbf{x}_{t+\Delta}$\texttt{}~\\
\texttt{Update time $t\mathrel{{+}{=}}\Delta$ }~\\
\texttt{Switch to MSM mode}
\item \texttt{Else (return event):}~\\
\texttt{Map to coordinates $\mathbf{c}_{t+\Delta}=\mathbf{x}_{t+\Delta}$}~\\
\texttt{Update time $t\mathrel{{+}{=}}\Delta$}
\end{itemize}
\end{enumerate}
\item \texttt{Else (MSM mode):}
\begin{enumerate}
\item \texttt{If $s_{t}\neq s_{t-\tau_{MSM}}$or previous mode $\neq$ MSM
mode:}
\begin{itemize}
\item \texttt{Sample next event ($\mathbf{x}_{t+\Delta}$,$\Delta$) from
}~\\
$p_{\mathrm{exit}}\left[s_{t}\rightarrow\mathbf{x}_{\mathrm{t}+\Delta};\,\Delta\right]$\texttt{. }
\item \texttt{If exit event:}~\\
\texttt{Map to coordinates $\mathbf{c}_{t+\Delta}=\mathbf{x}_{t+\Delta}$}~\\
\texttt{Update time $t\mathrel{{+}{=}}\Delta$}~\\
\texttt{Switch to RD mode}\texttt{\textcolor{blue}{{} }}\texttt{\textcolor{black}{and
break current loop iteration}}
\end{itemize}
\item \texttt{Propagate $s_{t}\rightarrow s_{t+\tau_{MSM}}$ using the MSM}
\item \texttt{Update time} $t\mathrel{{+}{=}}\tau_{MSM}$
\end{enumerate}
\end{enumerate}
There are additional issues in specific scheme implementations, such
as estimating the unknown conditional transfer probabilities, and
choosing the MSM discretization and $R$ such that the overall discretization
error is small, among others. These issues are non-trivial and could
potentially be tackled with different approaches. In order to quantify
the accuracy of a given approach, we quantify how well is our scheme
approximating the ground truth by comparing relevant macroscopic observables.
We present one possible implementation of the scheme in Sec. \ref{sec:MSM/RD-implementation}.

\section{An MSM/RD implementation for protein-ligand systems\label{sec:MSM/RD-implementation}}

\begin{figure*}
\centering

(a)\includegraphics[width=0.3\textwidth]{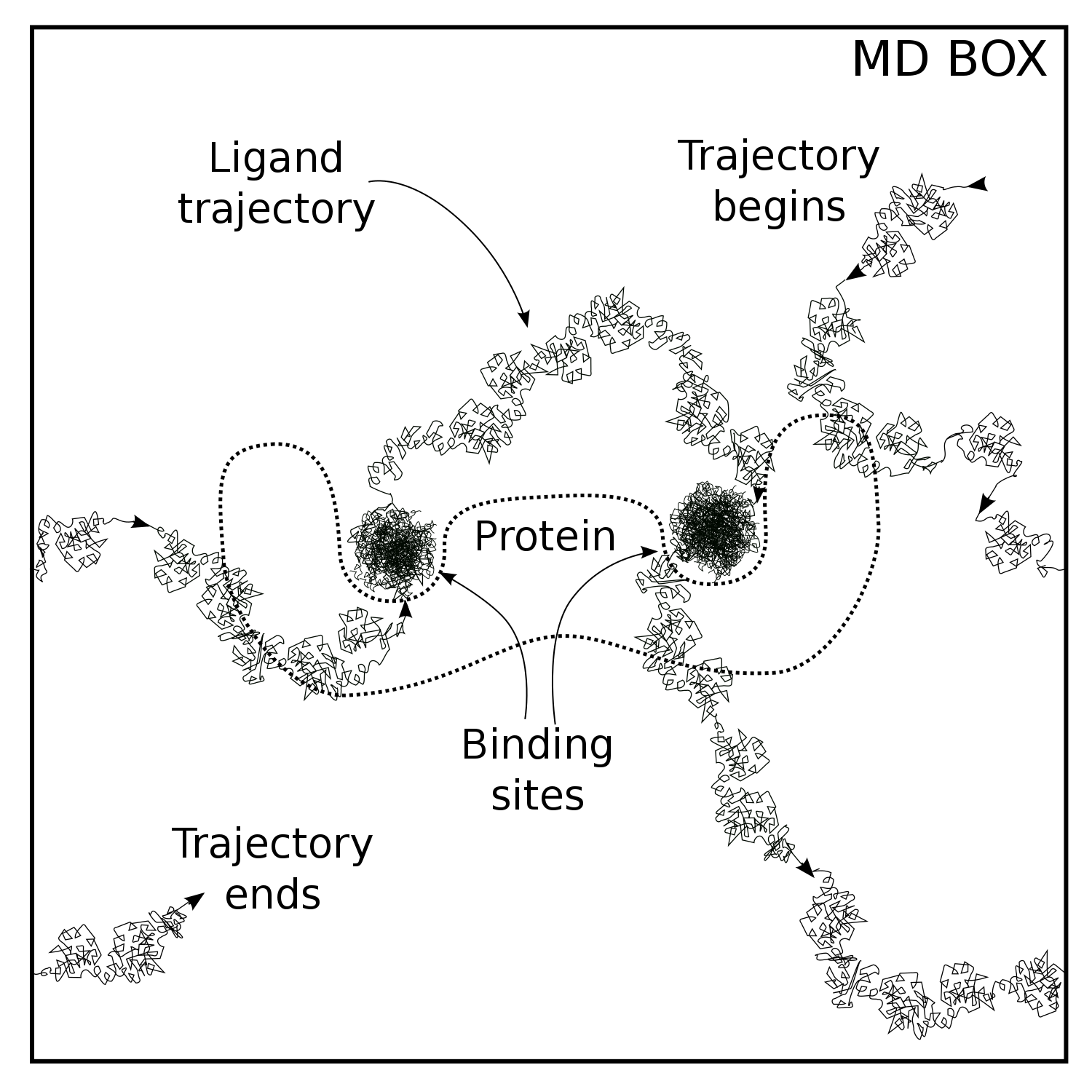}
(b)\includegraphics[width=0.3\textwidth]{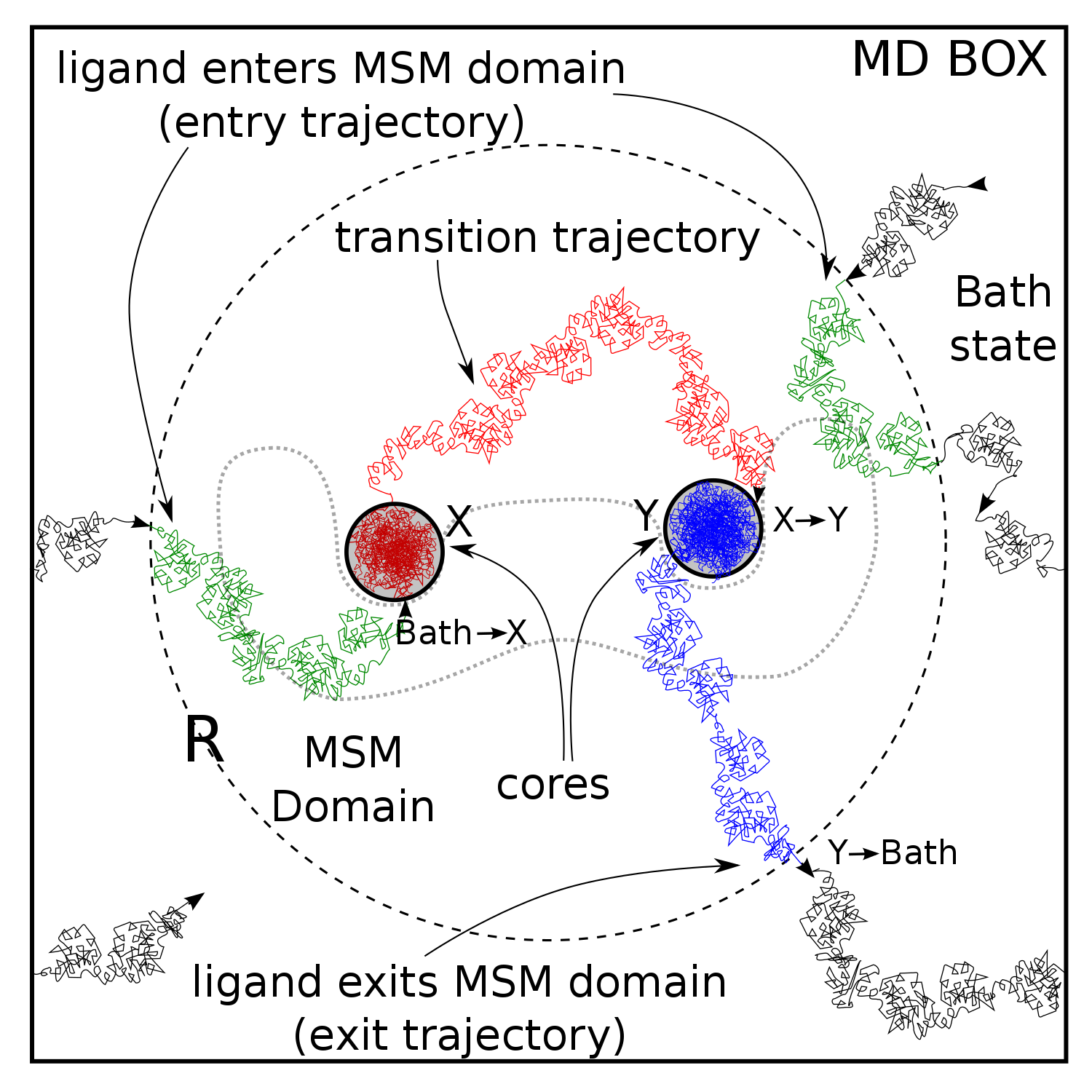}
(c) \includegraphics[width=0.3\textwidth]{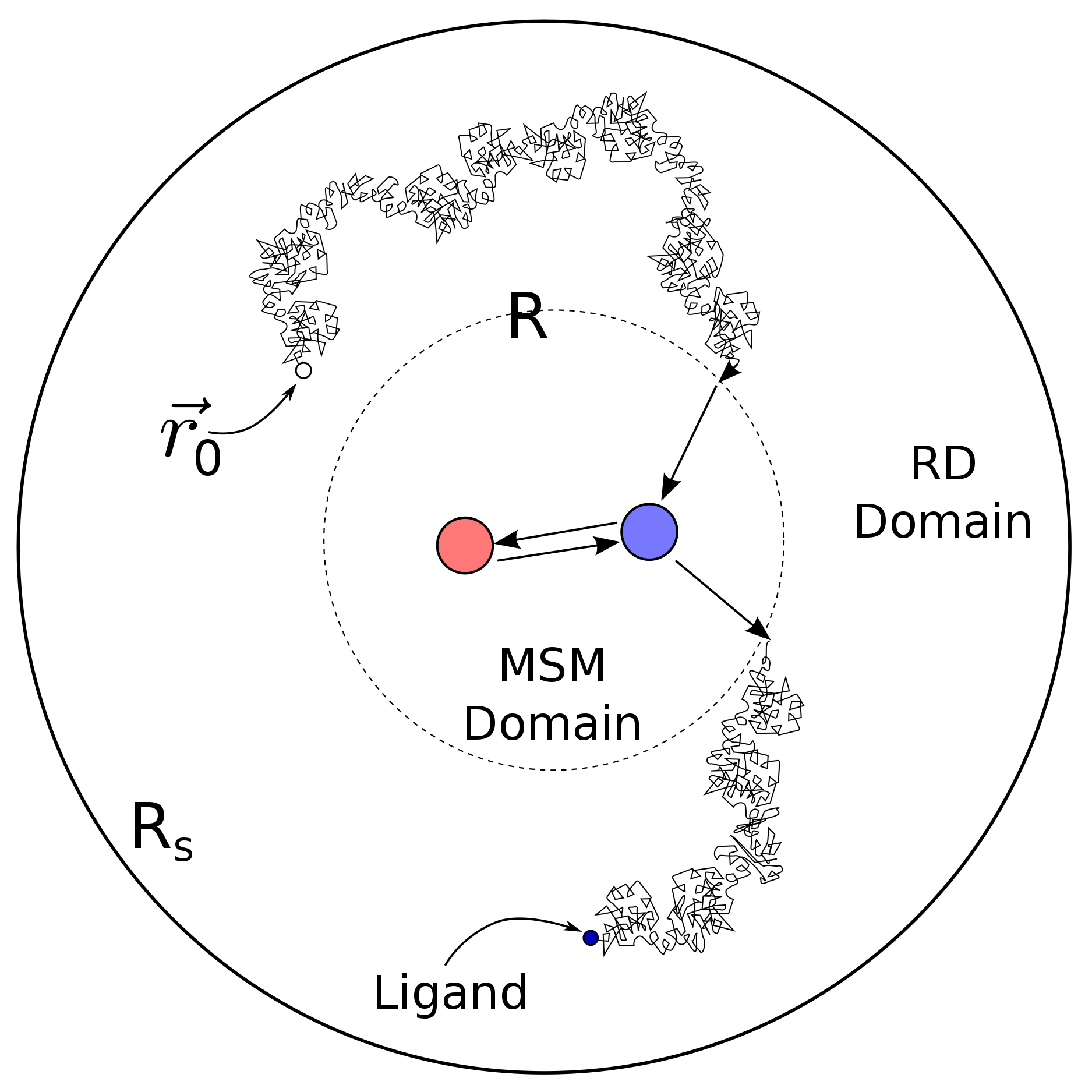}

\caption{Illustrations of a MD trajectory, its classification to extract the
relevant dynamics and the MSM/RD scheme.\textbf{ a)} Illustration
of a trajectory of a ligand in a MD simulation within a box with periodic
boundaries (small-scale simulation). Note that there are two metastable
regions, e.g. binding sites on a protein, where the ligand stays for
a longer time.\textbf{ b)} Illustration of truncation and classification
of the trajectory. The MSM domain is chosen so the interaction potential
is effectively zero outside this region (bath state); the cores $X$
and $Y$ are chosen to represent the metastable regions in phase space.
The truncated trajectories are classified into entry trajectories
(green), transition trajectories (red) and exit trajectories (blue),
which are used for the coupling in the MSM/RD scheme. In order to
obtain the MSM for the MSM/RD scheme, the system is also classified
into three states, the bath state and the two cores $X$ and $Y$;
it is also shown when the transition between these states occur along
a trajectory. \textbf{c)} Representation of the MSM/RD scheme. The
full trajectories from the MD-simulation are used to derive an MSM
to model the dynamics in the MSM domain. The entry and exit trajectories
from the MD-simulation are used to couple the Brownian dynamics in
the diffusion domain with the dynamics in the MSM domain. }

\label{fig:trajectoriesIllustration} 
\end{figure*}

Now we develop an implementation of the MSM/RD scheme for a special
class of systems: th\textcolor{black}{e binding of a small ligand
to a protein – a case that is relevant in the study of protein-drug
binding kinetics \citep{schuetz2017kinetics}. While the theory described
before is more general, implementations to more challenging systems
such as protein-protein interaction will be treated in future contributions.
We begin by considering the macromolecule $A$ fixed at the origin
with fixed orientation and the ligand $B$ freely diffusing around
it with an overall diffusion constant $D=D_{A}+D_{B}$. The macromolecule
has several possible binding sites given by some interaction po}tential.
In order to present the MSM/RD scheme in detail, we distinguish three
different simulations: 
\begin{enumerate}
\item \textbf{Reference simulation} (ground truth, if available): MD simulation
of $B$ and its interaction with $A$ in a large spherical domain
with radius $R_{s}$. Unfortunately, reference simulations of realistic
systems are in general not computationally feasible due to the time
and lengthscales of the simulation. Nonetheless, reference simulations
of simple systems are used to verify the MSM/RD scheme and validate
its use in more complex systems. 
\item \textbf{Small-scale simulation} (MD simulation): analogous to the
reference simulation with the difference that $B$ is constrained
to a small box with periodic boundary conditions, see Fig. \ref{fig:trajectoriesIllustration}a.
As the potential is negligible outside this box, the main interaction
dynamics are extracted from this simulation's data into an MSM. This
simulation is used to parametrize the MSM/RD model.
\item \textbf{MSM/RD simulation} (hybrid model): couples the MSM for short-range
interactions derived from the MD simulation 2. with a diffusion scheme
for the long-range, see Fig. \ref{fig:trajectoriesIllustration}c.
The goal of the scheme is to approximate the ground truth dynamics
given by the reference simulation 1. 
\end{enumerate}

\subsection{Estimation}

In order to parametrize the MSM/RD scheme, we need to estimate quantities
from the small-scale simulation 2. that characterize the state-to-state
dynamics and the coupling between the MSM and RD domain. The state
to state dynamics are estimated using an MSM, and the coupling is
given in terms of entry and exit events from the MSM domain. These
might happen on different timescales, so we would like to be free
from the fixed time-step the MSM requires to be well equilibrated.
Therefore, we use trajectory statistics for entry and exit events.

\subsubsection{MSM}

As a first step for the construction of the Markov model and MSM/RD
parametrization, we need to find a discrete representation of the
underlying data. In this work we use the core MSM approach \citep{schutte2011markov},
which requires the definition of cores as metastable regions of phase
space. Cores are given by spherical domains around the metastable
regions in the MD simulation and can be found using a clustering algorithm.
In the core MSM approach a discrete trajectory is constructed by assigning
the last visited state-index to each point in the trajectory. Note
the trajectory may leave the core of a given state and re-enter multiple
times without transitioning to other states. Using this discretization
technique we truncate the discrete trajectory into three types of
trajectories as shown in Fig. \ref{fig:trajectoriesIllustration}b:
$i)$ entry trajectories that start just inside the MSM domain and
either leave the domain next or hit a core inside the domain. $ii)$
transition trajectories that start in a state and hit another state
as next event and $iii)$ exit trajectories that start in a state
and leave the MSM domain as next event. These trajectories are used
to estimate the transfer densities and to parametrize the MSM/RD simulation.

The MSM for the short-range interactions is built using the full discrete
trajectories and the exit trajectories (Fig. \ref{fig:trajectoriesIllustration}a).
We follow the methods from \citep{PrinzEtAl_JCP10_MSM1} to estimate
a transition matrix $\mathbf{T}(\tau)$, where the entries are the
transition probabilities $T_{i,j}$ from state $i$ to $j$. Using
the discrete trajectories, we create count matrices $C_{i,j}^{\mathrm{full}}(\tau)$
from the complete data set and $C_{i,j}^{\mathrm{exit}}(\tau)$ from
the exit trajectories, which count all the transitions from state
$i$ to $j$ at a lag time $\tau$ observed in the respective datasets.
As the coupling between the MSM and RD domain is handled separately,
the MSM dynamics only accounts for transitions amongst the cores and
therefore the counts arising from exit trajectories have to be subtracted
\begin{align}
C_{i,j}(\tau)=C_{i,j}^{\mathrm{full}}(\tau)-\delta_{i,j}C_{i,j}^{\mathrm{exit}}(\tau),\label{eqn:countMatrix}
\end{align}
where $\delta_{i,j}$ denotes the Kronecker delta. We then use a maximum
likelihood estimator to obtain a transition matrix from the given
counts $C_{ij}$. Note that here we have chosen an irreversible estimator,
as we can no longer assume that detailed balance holds for this count
matrix.

\subsubsection{Entering the MSM domain\label{subsec:EnterMSM}}

 The protocol to enter the MSM domain from the RD domain is constructed
with the entry trajectories as defined above. It consists of generating
a list $L_{\mathrm{entry}}=\{\mathbf{c}_{\mathrm{entry}},\mathbf{x}_{\mathrm{end}},\Delta\}$
of all start coordinates $\mathbf{c}_{\mathrm{entry}}$(just inside
the MSM domain) and endpoints $\mathbf{x}_{\mathrm{end}}$ of entry
trajectories and their corresponding times $\Delta$. The endpoints
may be either MSM states or coordinates in the RD domain, see Fig.
\ref{fig:trajectoriesIllustration}b. The ensemble of trajectories
in this list estimates the conditional transfer probability \texttt{$p_{\mathrm{entry}}\left[\mathbf{c}_{t}\rightarrow s_{t+\Delta};\,\Delta\right]$}
(Sec. \ref{subsec:MSM/RD-coupled}) for several times $\Delta$. In
the MSM/RD simulation samples are drawn from this list of entry points.

\subsubsection{Exiting the MSM domain\label{subsec:ExitMSM}}

For each state $\mathbf{s}$ of the MSM, we collect all exit and transition
trajectories and save their end coordinate or state along with their
respective exit time in the lists $L_{\mathrm{exit},s}=\{\mathbf{c}_{exit},\Delta\}$
and $L_{\mathrm{trans},s}=\{s_{\mathrm{trans}},\Delta\}$. The ensemble
of trajectories in these list estimates the conditional transfer probability
\texttt{$p_{\mathrm{exit}}\left[s_{t}\rightarrow\mathbf{c}_{\mathrm{t}+\Delta};\,\Delta\right]$
}(Sec. \ref{subsec:MSM/RD-coupled}) for several times $\Delta$.
The probability of an exit event $P_{\mathrm{exit},s}$ is simply
estimated as the ratio of exiting trajectories over the total numbers
of trajectories,

\begin{align}
P_{\mathrm{exit},s}=\frac{\#\text{\text{ of trajectories in} }L_{\mathrm{exit},s}}{\#\text{\text{ of trajectories in} }L_{\mathrm{exit},s}\text{ and }L_{\mathrm{trans},s}}.\label{eqn:exitProbability-1}
\end{align}

The MSM/RD scheme for this implementation is based on the scheme from
Sec. \ref{subsec:MSM/RD-coupled} and is shown in the Appendix.

\subsection{Verification of the MSM/RD scheme}

In order to verify the MSM/RD scheme, we use systems where a reference
simulation is available. We verify the internal dynamics by comparing
the first passage times (FPTs) distributions and mean first passage
times (MFPTs) for each pair of metastable states within region $I$
between the MSM/RD and reference simulations. We estimate the ground
truth MFPTs by computing the FPTs $t_{i,j}^{\textrm{ref}}$, where
the initial conditions are chosen as the minima $\mu_{i}$ and the
system is propagated following the reference simulation until hitting
state $j$ (conditioned on not leaving the MSM domain). For the MSM/RD
scheme, we compute the FPTs $t_{i,j}^{\textrm{MSM}}$ by placing the
particle in state $i$ and propagating the system following the MSM/RD
scheme until state $j$ is hit. If the particle exits the MSM domain
before reaching state $j$, the trajectory is not taken into account.
When a sufficiently large sample is generated, we can estimate the
distributions of FPTs by histograms. The MFPTs are estimated as $\tau_{ij}^{\textrm{ref}}$=$\overline{t_{ij}^{\textrm{ref}}}$
and $\tau_{ij}^{\textrm{MSM}}$=$\overline{t_{ij}^{\textrm{MSM}}}$,
respectively. The MFPT relative error between the MSM/RD and the reference
simulations is estimated as 
\begin{align}
(E_{\text{rel}})_{\mathrm{ij}}=\frac{\tau_{ij}^{\textrm{ref}}-\tau_{ij}^{\mathrm{MSM}\mathscr{}}}{\tau_{ij}^{\textrm{ref}}}.\label{eqn:relativeError}
\end{align}
In order to verify the coupling between the RD and MSM domain, we
also estimate and compare the unbinding rate, binding rate and equilibrium
constant. The two latter are calculated for different particle concentrations
$c$ by fixing the radius $R_{s}$ of the simulation domain such that
$c=1/V_{\text{RD}},$with $V_{\mathrm{RD}}$ the volume of the RD
domain.

\section{Results\label{sec:results}}

In this section, we implement the MSM/RD scheme from Sec. \ref{sec:MSM/RD-implementation}
in two systems. The first is a simple model of a ligand diffusing
in a potential landscape, which is used to verify that the MSM/RD
scheme reproduces the correct dynamics. The second corresponds to
a more realistic MD system, where we study the binding of carbon monoxide
to myoglobin.

\subsection{Ligand diffusion in potential landscape\label{subsec:3D-diff-pot}}

We implement the MSM/RD scheme in a simple model, where the reference
simulation is available. The model consists of a ligand $B$ under
over-damped Langevin dynamics in a three-dimensional potential landscape
\begin{equation}
\frac{d\mathbf{x}(t)}{dt}=-\frac{1}{\gamma}\nabla U(\mathbf{x})+\sqrt{2D}\boldsymbol{\xi}(t),\label{eq:overdLangevin}
\end{equation}
with $U$ the interaction potential with some macromolecule $A$ fixed
at the origin, $\gamma$ the damping, and each component of the noise
satisfies $E[\xi_{i}(t)]=0$ and $E[\xi_{i}(t)\xi_{j}(s)]=\delta_{ij}\delta(t-s)$
with $D=k_{B}T/\gamma$ the diffusion coefficient. A trajectory density
plot of the potential landscape chosen is shown in Fig. \ref{fig:diff3Dpot}a,
and it consists of nine Gaussians with different depths and widths
\begin{align}
U(\mathbf{r})=-\sum_{i=1}^{9}s_{i}\mathcal{N}(\mathbf{\boldsymbol{\mu}}_{i},\mathbf{\boldsymbol{\Sigma}}_{i}),\label{eqn:potentialLandscape}
\end{align}
where $\mathcal{N}(\boldsymbol{\mu}_{i},\boldsymbol{\Sigma}_{i})$
denotes a Gaussian centered at minimum $\boldsymbol{\mu}_{i}$ with
covariance matrix $\boldsymbol{\Sigma}_{i}$, $s_{i}$ denotes a scale
factor. The small-scale simulation consists of Euler-Maruyama numerical
realizations of Eq. (\ref{eq:overdLangevin}) under this potential
constrained to a box with an edge length of $6$ units with periodic
boundary conditions. The reference simulation is analogous to the
small-scale simulation with the difference that it uses a larger spherical
domain with reflective boundary conditions at a range of radii corresponding
to simulations at different ligand concentrations. 

\subsubsection{Parametrization of the MSM/RD scheme}

We use a radiu\textcolor{black}{s of $R=2.5\:\mathrm{nm}$ for the
MSM domain ($I$ region) since outside this domain the potential (Eq.
\ref{eqn:potentialLandscape}) is essentially zero. We generate $120$
small-scale simulation trajectories, each with a length of $10^{7}$
steps, a time-step of $\Delta t=10^{-4}\:\mathrm{ns}$, and sampled
every tenth step. This results in a total simulation time of $t=1.2\cdot10^{5}\:\mathrm{ns}$. }

\begin{figure*}
\centering

(a) \includegraphics[width=0.3\textwidth]{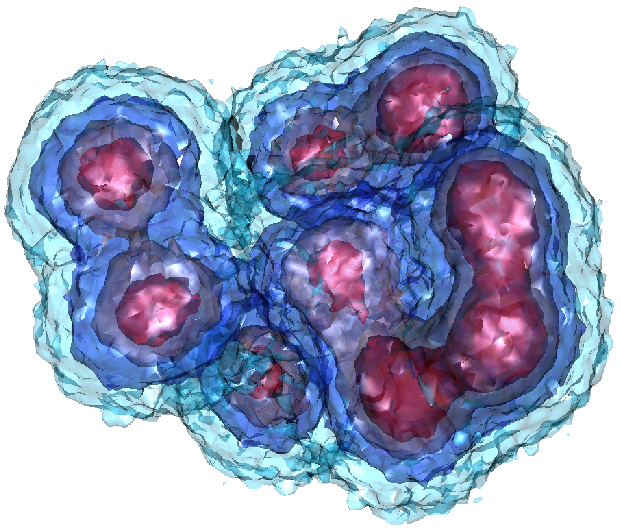}
(b)\includegraphics[width=0.3\textwidth]{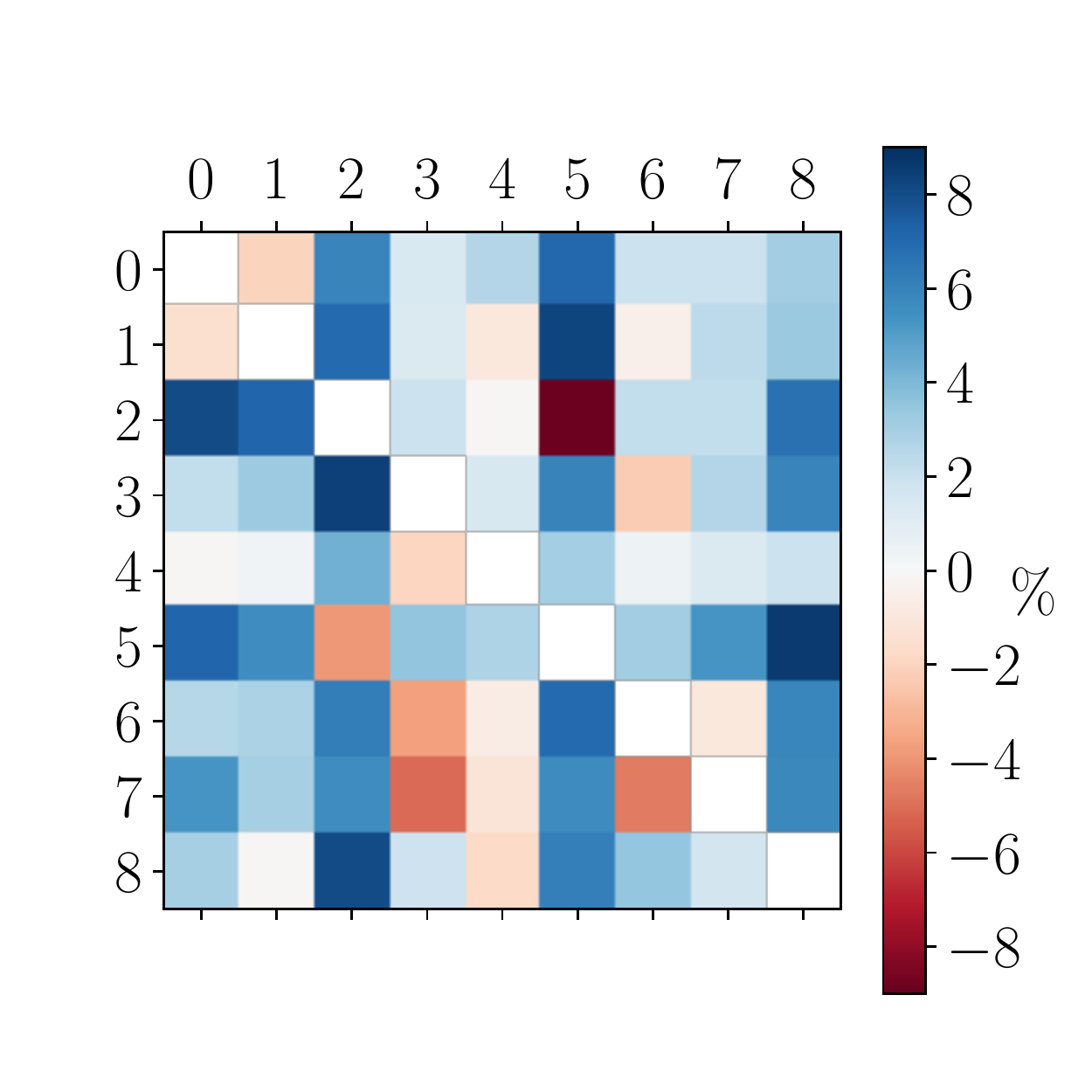}
\textcolor{blue}{(c)}\includegraphics[width=0.3\textwidth]{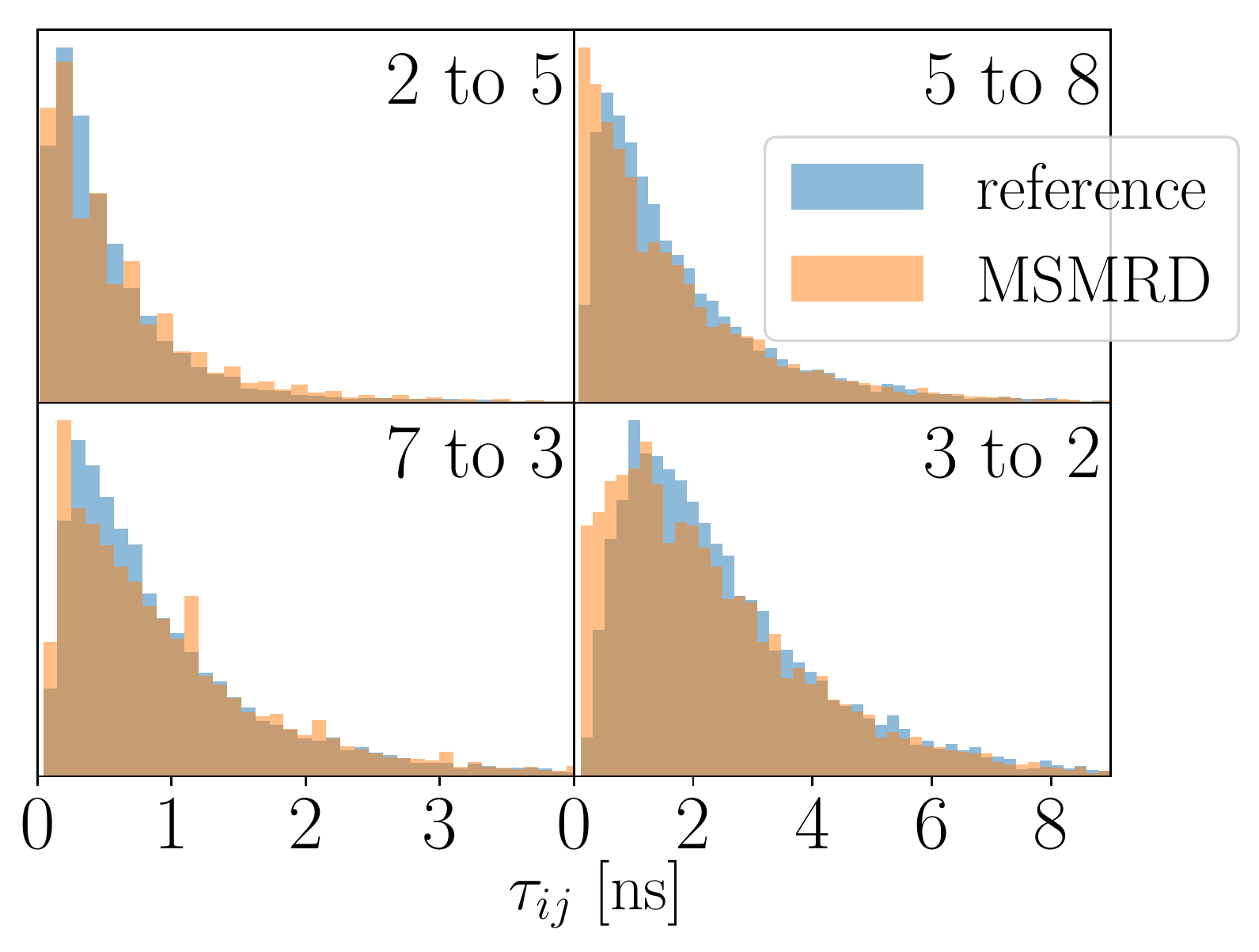}

\textcolor{blue}{(d)} \includegraphics[width=0.3\textwidth]{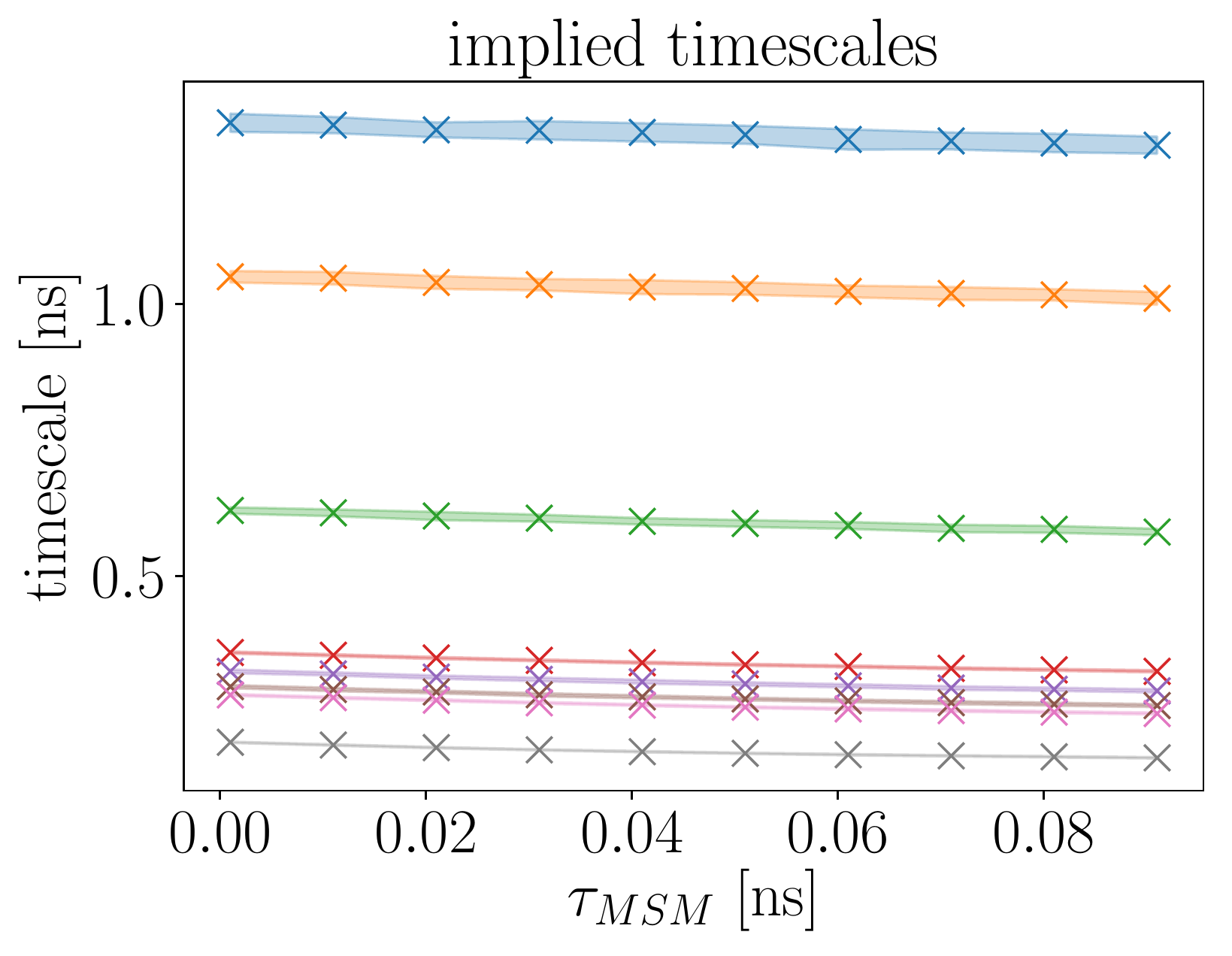}
\textcolor{blue}{(e)}\includegraphics[width=0.3\textwidth]{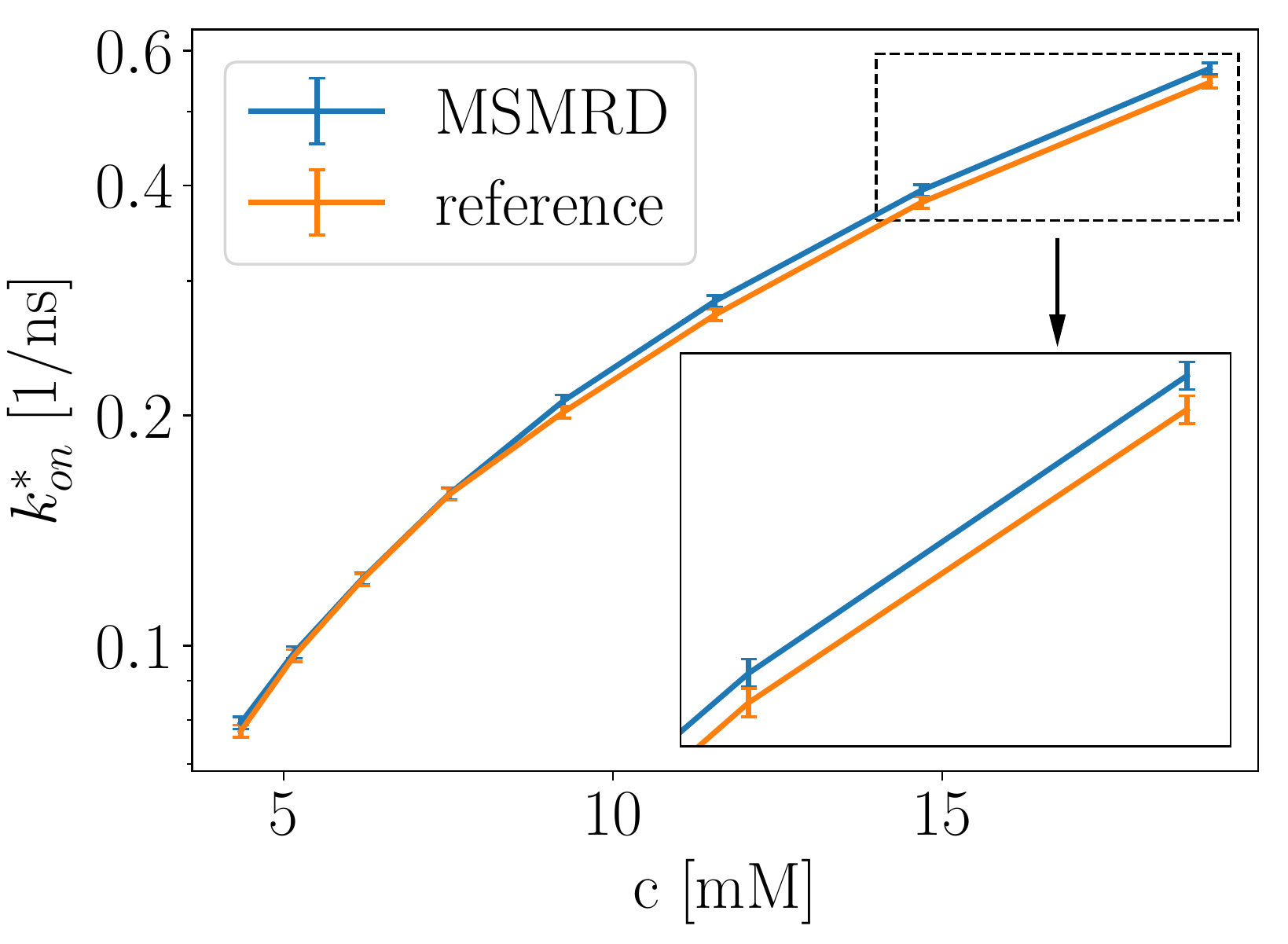}
\textcolor{blue}{(f)}\includegraphics[width=0.3\textwidth]{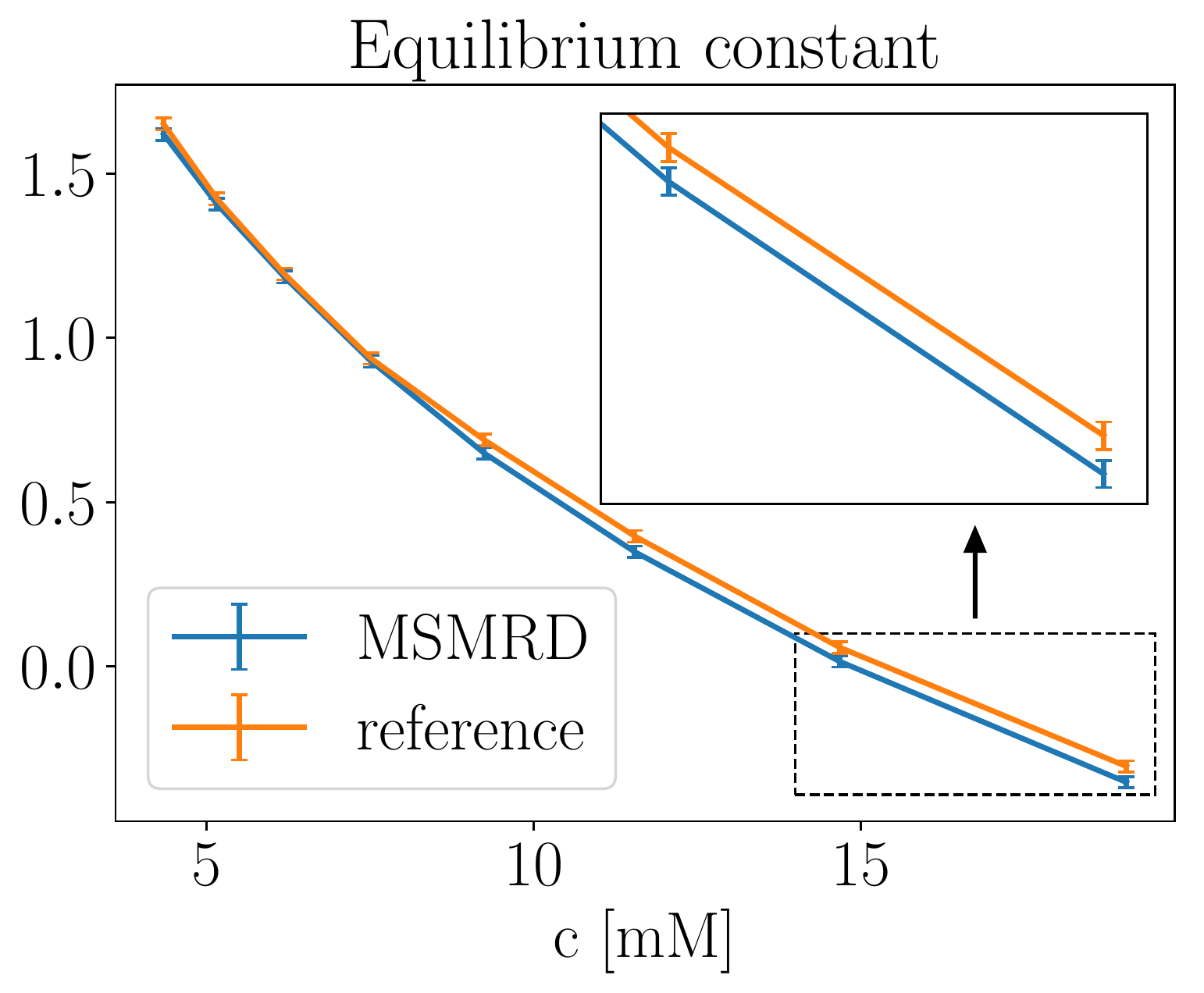}

\caption{Visualization and verification results for the simple model of ligand
diffusing in a potential landscape.\textbf{ (a)} Density plot of the
position of the ligand in the three dimensional potential. Red indicates
regions of higher density while blue indicates regions of lower density.
\textbf{(b)} Relative error of MFPTs conditioned on not leaving the
MSM domain between the MSM/RD and the reference simulation. \textbf{(c)
}Comparison of first passage times distribution histograms for the
transitions with the highest error in (b). The left pane corresponds
to transitions with negative relative error, and the right pane to
transitions with positive relative error. \textbf{(d)} Implied timescales
of the MSM. The shaded area represents the standard deviation of the
bootstrapping sample. We observe well converged timescales for all
considered lag times.\textbf{ (e)} The rate $k_{on}$ as function
of the concentration of the system for the MSM/RD and reference simulations.
\textbf{(f)} Same plot as e) but for the logarithm of the equilibrium
constant $\log(K_{eq})$. The error bars in (e) and (f) represent
the 95\% confidence interval using a bootstrapping approach.}

\label{fig:diff3Dpot}
\end{figure*}

\textcolor{black}{The cores are defined as spheres with radius $0.2\:\mathrm{nm}$
around the minima $\boldsymbol{\mu}_{i}$, and the count matrix of
transition between cores is generated from the trajectories following
Eq. (\ref{eqn:countMatrix}). A maximum likelihood estimator (implemented
in PyEMMA \citep{SchererEtAl_JCTC15_EMMA2}) is then applied to the
count matrix to yield the MSM. From the trajectories, we also generate
the lists $L_{\mathrm{entry}}$, $L_{\mathrm{exit},s}$, $L_{\mathrm{trans},s}$
and $P_{\mathrm{exit},s}$, introduced in Secs. \ref{subsec:EnterMSM}
and \ref{subsec:ExitMSM}. We then estimate the timescales of the
eigenmodes for different MSM lag times to test how well the underlying
process is estimated by the MSM. The timescales have small variations
for different lag times (Fig. \ref{fig:diff3Dpot}d), which means
the system can be considered Markovian for all lag times. However,
we have to be careful not to choose the lag time too large, such that
relevant fast timescales are neglected resulting in significant errors.
For all further analyses, we consider a lag time of $\tau_{MSM}=500\Delta t=0.05\:\mathrm{ns}$
to be a}n optimal compromise.

\subsubsection{Comparison of dynamic properties}

In order to compute the binding rate, we calculate the first passage
time from a uniforml\textcolor{black}{y sampled location close to
the boundary $r=R_{S}-\delta$ to any MSM state. We choose $\delta=0.05\:\mathrm{nm}$
and use $10^{4}$ simulations to average and estimate the $\text{MFPT}_{\mathrm{on}}$,
from which we calculate the binding rate as $k_{\mathrm{on}}^{*}=1/\text{MFPT}_{\mathrm{on}}$.
This procedure is performed for both the MSM/RD and the reference
simulation, and we observe excellent agreement between the two (Fig.
\ref{fig:diff3Dpot}e).}

\textcolor{black}{For the unbinding rate, we consider the inverse
process by starting in an MSM state and propagating the dynamics until
crossing a boundary defined by a sphere with radius $2.7\:\mathrm{nm}>R$.
We obtain a reference value of $0.402_{0.400}^{0.404}\:\mathrm{ns^{-1}}$
(Sub- and superscript indicate lower and upper bound of the $95\%$
precentile) and an MSM/RD simulation value of $k_{\mathrm{off}}=0.400_{0.398}^{0.402}\:\mathrm{ns^{-1}}$.
We further compute the logarithm of the equilibrium constant $\log(K_{\mathrm{eq}})=\log(k_{\mathrm{off}}/k_{\mathrm{on}}^{*})$
for both models and for the chosen values of concentrations, resulting
in accurate reproduction of the reference values by the MSM/RD scheme(Fig.
\ref{fig:diff3Dpot}f). Thus we verify that the coupling between the
MSM domain and the RD domain works consistently in the MSM/RD simulation
scheme.}

\textcolor{black}{Next, we want to ensure that also the dynamics between
the states inside the MSM are reproduced to a high accuracy. We compare
MFPTs between all pairs of states conditioned on not leaving the MSM
domain. In the reference simulation this is done by placing the particle
at position $\boldsymbol{\mu}_{i}$ and propagating the system until
state $j$ is reached. If the particle leaves the MSM domain before
reaching state $j$, this trajectory is discarded. For the MSM/RD
simulation, we simply start in state $i$ and propagate until state
$j$ is hit, while discarding trajectories that leave the MSM domain.
This procedure is repeated until $10^{4}$ successful trajectories
are found for both simulations, w}hich are averaged to obtain the
MFPTs. The relative errors are calculated with Eq. (\ref{eqn:relativeError});
all relative errors are below $9\%$ (Fig. \ref{fig:diff3Dpot}b).
We further observe that negative errors arise for state pairs that
are close together and thus have short passage times. For these transitions,
we tend to overestimate the MFPT in the MSM/RD simulation as short
processes are truncated in the MSM estimation. Moreover, we observe
that the highest positive errors arise for transitions which are far
apart. These are the hardest to sample since for these transitions
there are a very high number of possible long and non-direct transition
trajectories, which are less likely to be observed . We chose the
four transitions with the highest relative error and compared their
FPTs distribution histograms (Fig. \ref{fig:diff3Dpot}c). Even though
these transitions have the highest errors, we observe the distributions
match well. Therefore, we verify MSM/RD scheme also describes the
internal dynamics accurately.

\subsection{Binding of CO to myoglobin\label{subsec:BindingCO}}

As an application of the MSM/RD scheme, we study the binding of carbon
monoxide (CO) to myoglobin. Myoglobin is a globular protein which
is responsible for the transport of oxygen in muscle tissue. The binding
process of CO to myoglobin has recently be\textcolor{black}{en studied
by de Sancho et al. \citep{deSancho2015identification}, whose data
we use to parametrize the MSM/RD scheme. The dataset consist of MD
trajectories of $20$ CO molecules and one myoglobin protein for a
total simulation time of $500\:\mathrm{ns}$. The MD simulation is
confined to a periodic box with edge length of $5\:\mathrm{nm}$.
Despite the fact that only one CO molecule can reside in the binding
pocket, the error of treating $20$ CO molecules as being statistically
independent is small within statistical uncertainty (see  \citep{deSancho2015identification}
for details). We therefore extract$20$ independent CO trajectories,
effectively  increasing the total simulation time to $10\:\mu\mathrm{s}$.}

\subsubsection{Parametrization of MSM/RD scheme}

\begin{figure}
\centering

(a) \includegraphics[width=0.8\columnwidth]{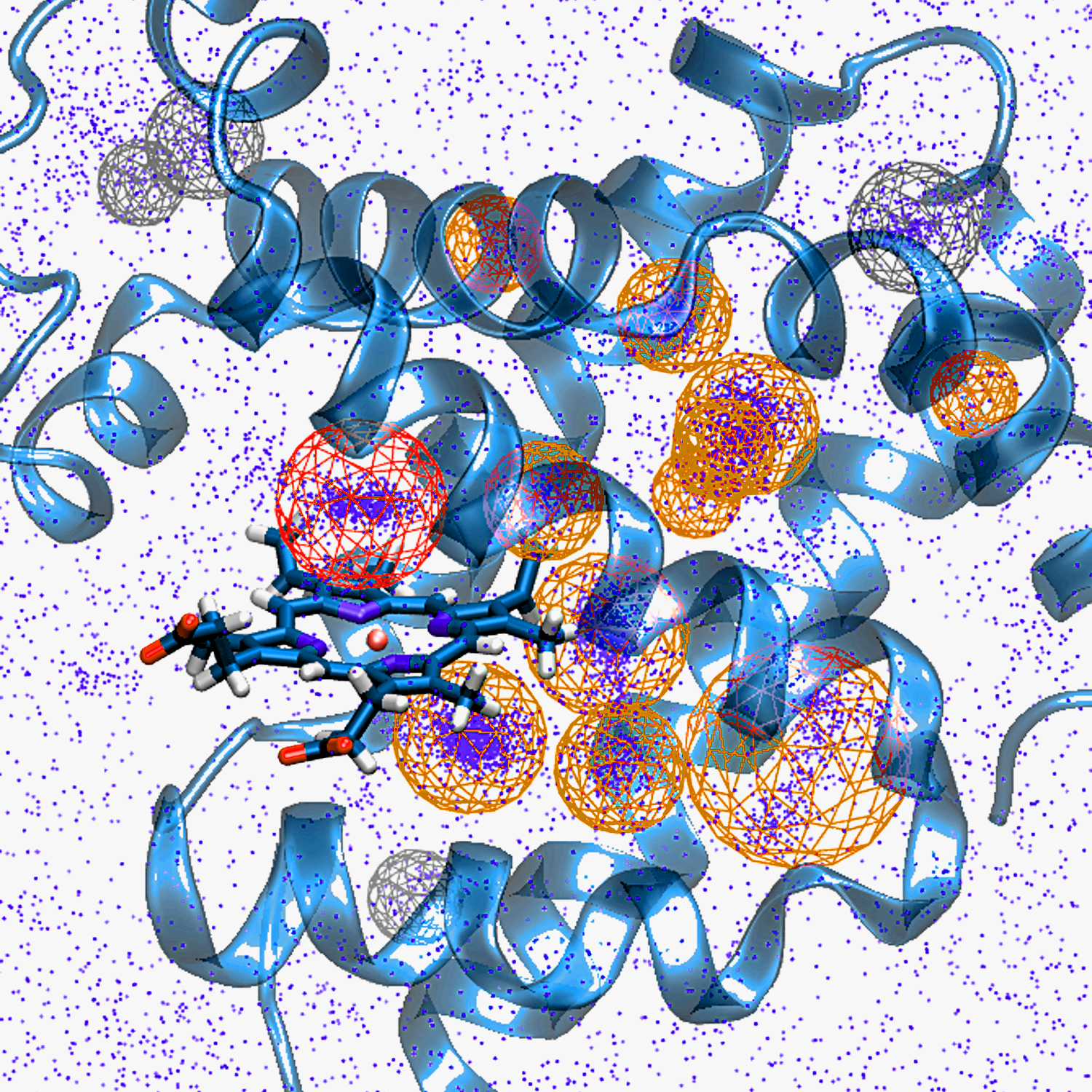}

(b)\includegraphics[width=0.8\columnwidth]{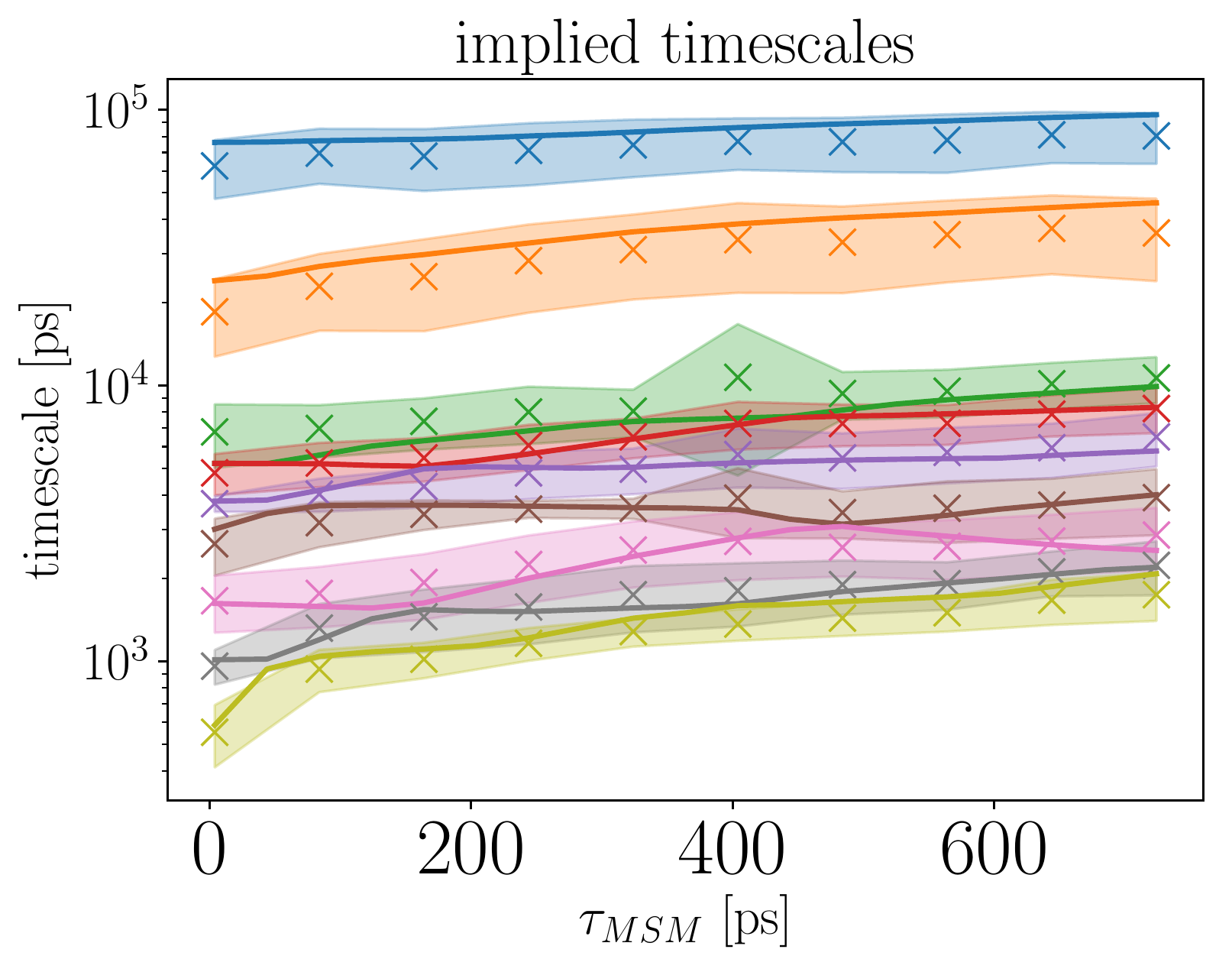}

(c)\includegraphics[width=0.8\columnwidth]{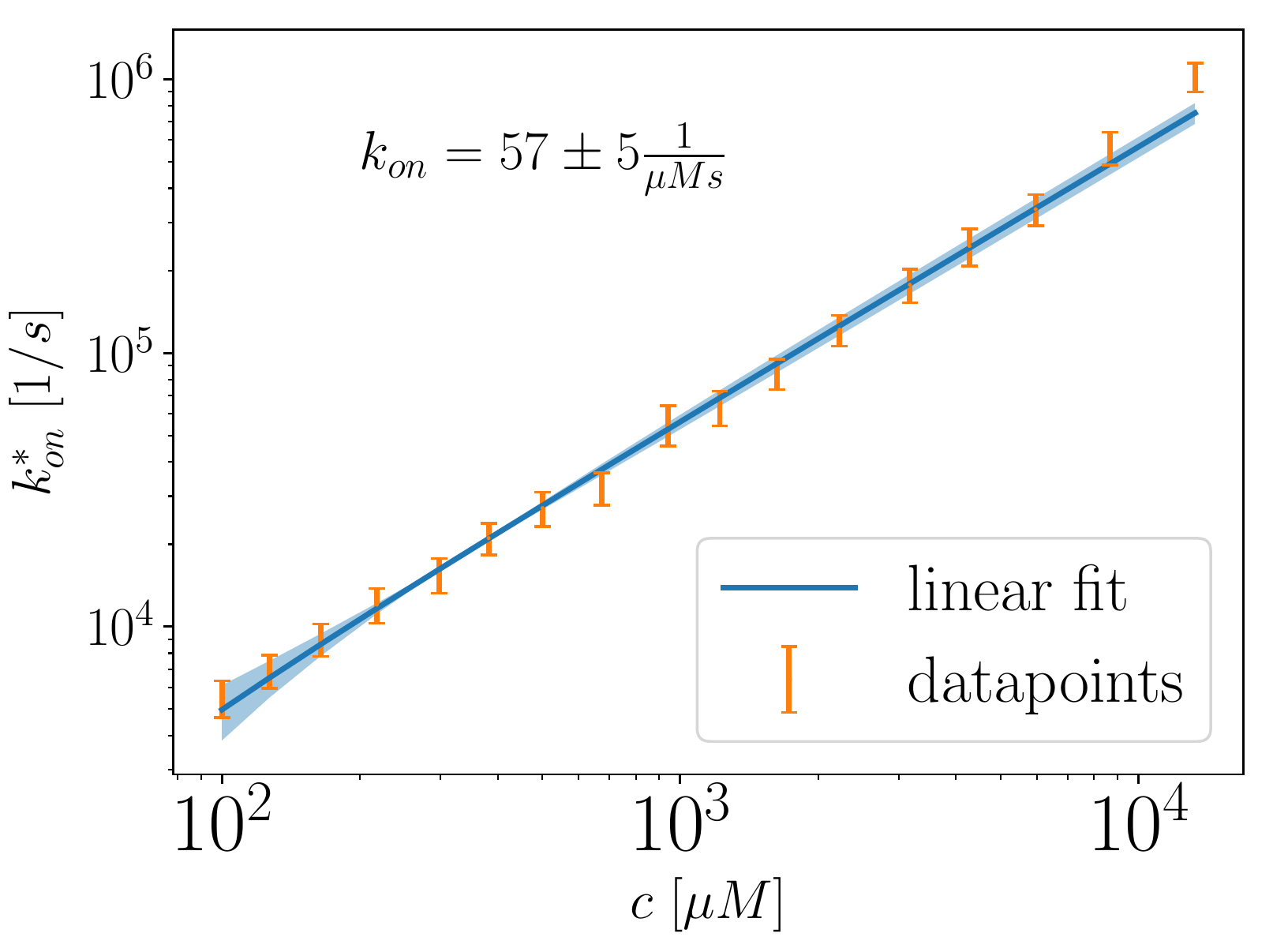}

\caption{\textcolor{black}{Discretization and results of the CO-myoglobin system.}\textbf{\textcolor{black}{{}
(a)}}\textcolor{black}{{} Definition of the cores (wire frame spheres)
within the myoglobin. The red sphere indicates the bound state. The
gray spheres correspond to the states that were not in the connected
set and therefore discarded. The blue dots are positions of the CO
molecules for every 50th frame in the vicinity of the protein. }\textbf{\textcolor{black}{b)}}\textcolor{black}{{}
Implied timescales of the dynamics of the CO myoglobin system. The
datapoints and shaded area denote the sample mean and standard deviation
of the bootstrapping sample over the trajectories: from the 20 given
trajectories we resample 20 with replacement. Over this sample we
run our discretization process which returns a sample of timescales.
The trajectory-samples which are not ergodic or do not lead to a connected
count matrix are considered invalid and discarded. Solid lines are
found using the full dataset. }\textbf{\textcolor{black}{c)}}\textcolor{black}{{}
Reaction rate as estimated from multiple simulations at different
concentrations.}}

\label{fig:Myoglobin}
\end{figure}

In order to parametrize the scheme, all frames are first \textcolor{black}{aligned
using the $C_{\alpha}$ atoms of the myoglobin as reference. On the
aligned data, we run the density-based spatial clustering of applications
with noise algorithm (DBSCAN) \citep{ester1996density}, which finds
a total of 16 metastable regions/cores. The positions and size of
the cores are shown in Fig. \ref{fig:Myoglobin}a, where it can be
observed that the algorithm correctly identifies regions of high ligand
density, including the myoglobin bound state indicated in red. The
radius of the spherical cores is the radius at which $80\:\%$ of
the datapoints that were assigned to the respective state are inside
the core. Four states are discarded as they are not part of the largest
connected set. As the simulation box had been set up to just contain
the protein and a $1\:\mathrm{nm}$ solvent layer, we choose the largest
MSM domain that still fits inside the box $(R=2.5\:\mathrm{nm)}$.
Analogous to the previous example, we follow Sec. \ref{sec:MSM/RD-implementation}
to estimate an MSM for the close-range dynamics and generate $L_{\mathrm{entry}}$,
$L_{\mathrm{exit},s}$, $L_{\mathrm{trans},s}$ and $P_{\mathrm{exit},s}$
to couple the dynamics in the two domains. }

\textcolor{black}{We compute the implied timescales for the MSM and
choose a lag time of $150\:\mathrm{ps}$ where timescales are sufficiently
converged (Fig. \ref{fig:Myoglobin}b). The diffusion constant is
computed using the mean squared displacement (MSD) of the parts of
the CO trajectories that are far from the protein, with $D=\Delta\text{MSD}(t)/6\Delta t$.
We find a diffusion constant of $\ensuremath{D_{\text{CO}}=2.5\:\mathrm{nm}^{2}\mathrm{ns}^{-1}}$,
which is comparable to the experimental value which is in the range
of $D_{\text{CO}}=2.03\:\mathrm{nm}^{2}\mathrm{ns}^{-1}$ (at $20\:C\text{\textdegree}$)
to $D_{\text{CO}}=2.43\:\mathrm{nm}^{2}\mathrm{ns}^{-1}$ (at $30\:C\text{\textdegree}$)
\citep{wise1968diffusion}.}

\subsubsection{Comparison of dynamic properties}

\textcolor{black}{As in the previous example, we compute the binding
rate by sampling positions sampled uniformly in the RD domain and
simulating the MSM/RD model until it reaches the bound state. For
each concentration, 200 trajectories are run to estimate the binding
rate $k_{\mathrm{on}}^{*}$. These rates are plotted against the concentration
and shown in Fig. \ref{fig:Myoglobin}c. The reaction rate $k_{\mathrm{on}}=57_{52}^{62}\:\mu\mathrm{M}^{-1}\mathrm{s}^{-1}$
is obtained as the slope of the linear fit. For the unbinding rate,
we start simulations in the bound state and collect MFPTs for leaving
the MSM domain; we find a rate of $k_{\mathrm{off}}=19.0_{18.8}^{19.2}\:\mu\mathrm{s}^{-1}$.
The resulting equilibrium constant $K_{\mathrm{eq}}=k_{\mathrm{on}}/k_{\mathrm{off}}=3.0_{2.7}^{3.3}\:\mathrm{M}^{-1}$
is similar to $3.6\,\mathrm{M}^{-1}$ found by de Sancho et al. \citep{deSancho2015identification},
both of which are close to the experimental value of $2.2\:\mathrm{M}^{-1}$\citep{carver1990analysis}
(see Tab. \ref{tab:Myoglobin_kinetics} for comparison). The binding
rate and unbinding rate found by de Sancho et al. \citep{deSancho2015identification},
although yielding a similar equilibrium constant, are both nearly
an order of magnitude faster than the ones obtained with MSM/RD (Tab.
\ref{tab:Myoglobin_kinetics}). The first indication that the present
rates are an improved estimate is the fact that the kinetics (both
the MSM relaxation timescales and $k_{\mathrm{on}}$) are independent
of the lag time (Fig. \ref{fig:Myoglobin}b, c). }

\textcolor{black}{To validate that the MSM/RD estimates of $k_{\mathrm{off}}$
and $k_{\mathrm{on}}$ have been estimated without significant bias,
it must be shown that they are statistically consistent with the ground
truth (in this case a sufficiently large and sufficiently long MD
simulation). Here, $k_{\mathrm{off}}$ can be estimated directly by
counting the frequency of ligand dissociation events from the binding
pocket in the underlying MD simulations. Since there are not sufficient
full dissociation pathways from the bound to the dissociated states
in the MD data in order to make a statistically relevant comparison,
we obtain a more precise estimate by computing the MFPT using an MSM
directly constructed from the original MD data with the same discretization
as used in the MSM/RD model. This resulted in a reference estimate
of $23.4_{11.6}^{46.6}\:\mu\mathrm{s}^{-1}$ (95\% percentile computed
with 1000 bootstrap samples), which is consistent with the MSM/RD
estimate (Tab. \ref{tab:Myoglobin_kinetics}).}

\textcolor{black}{Unfortunately, this method is not as accurate for
the binding rate $k_{\mathrm{on}}$, which is notoriously difficult
to estimate from small MD simulation boxes, where the length of trajectory
segments in which the ligand stays in the dissociated state without
touching the protein or crossing the periodic boundary are short compared
to lagtimes $\tau$ used in an MSM approach, resulting in biased estimates
\citep{PlattnerEtAl_NatChem17_BarBar}. Therefore, we performed another
Myoglobin MD simulation in an eightfold larger periodic box (edge
length $10\,\mathrm{nm}$) with the same CO concentration as in the
small MD simulation (resulting in $160$ CO molecules) for a total
simulation time of $405\:\mathrm{ns}$. For this data, a direct MSM
estimate of the binding rate yields $74.7_{29.9}^{130.9}\:\mu\mathrm{M}^{-1}\mathrm{s}^{-1}$
(95\% percentile computed with 1000 bootstrap samples). As a result,
the MSM/RD binding and dissociation rates are consistent with standard
estimates computed directly from MD simulation, and the MSM/RD modeling
error can be concluded to be statistically insignificant.}

\textcolor{black}{Given the consistency of the model,  we also compare
the results to experimental measurements, which is essentially a test
of the MD model (e.g. force field, thermostat, integrator). These
are yet a factor 4-5 slower than our estimates ($k_{\mathrm{on}}=12\:\mu\mathrm{M}^{-1}\mathrm{s}^{-1}$
and $k_{\mathrm{off}}=5.3\:\mu\mathrm{s}^{-1}$ found in \citep{carver1990analysis}),
confirming that the major part of the difference between the estimates
in \citep{deSancho2015identification} and theexperimental values
could be removed by the fact that MSM/RD is a significantly more accurate
model of the binding kinetics. }

\begin{table*}
\begin{tabular}{|c|>{\centering}p{2cm}|>{\centering}p{3cm}|c|>{\centering}p{2cm}|>{\centering}p{2cm}|}
\hline 
 & MSM/RD  & Reference (approx. ground truth) & MSM in \textcolor{black}{\citep{deSancho2015identification}} & Experiment \textcolor{black}{\citep{carver1990analysis}} & Unit\tabularnewline
\hline 
\hline 
$k_{\mathrm{on}}$ & $57.0_{52.0}^{62.0}$ & $74.7_{27.9}^{130.9}$ & 647 & $12$ & $\mathrm{M}^{-1}\mu\mathrm{s}^{-1}$\tabularnewline
\hline 
$k_{\mathrm{off}}$ & $19.0_{18.8}^{19.2}$ & $23.4_{11.6}^{46.6}$ & 179 & $5.3$ & $\mu s^{-1}$\tabularnewline
\hline 
$K_{\mathrm{eq}}$ & $3.0_{2.7}^{3.3}$ & $3.19_{2.6}^{3.8}$ & $3.6$ & $2.2$ & $\mathrm{M}^{-1}$\tabularnewline
\hline 
\end{tabular}

\caption{\textcolor{black}{\label{tab:Myoglobin_kinetics}Rates and equilibrium
constants for Myglobin-CO estimated from different methods. The reference
values approximate the ground truth by conducting a standard MSM-based
MFTP estimate from the MD simulation (for $k_{\mathrm{on}}$ a larger
simulation box was used to allow for a generous definition of the
dissociated state).}}
\end{table*}

\section{Conclusion\label{sec:Conclusion}}

We introduced and developed the MSM/RD scheme, which couples MD-derived
MSMs with RD simulations. We showed an implementation for protein-ligand
systems and applied it to two simple systems. The main advantage of
the algorithm is that it can simulate large time- and lengthscales
while conserving molecular resolution and computational efficiency.
This is achieved by extracting the characteristic features of the
dynamics fro\textcolor{black}{m several short MD simulations into
an MSM, which can produce new data with great accuracy and at a much
faster rate than the original MD simulations. This is a clear advantage
in comparison to previous works, like \citep{vijaykumar2015combining,VijaykumarEtAl_Arxiv16_AnisotropicMultiscaleGFRD},
since it does not require running MD simulations every time two particles
are close to each other. It can further yield more accurate binding
rates than traditional MSM methods by extending the diffusion domain
available, lessening the periodic boundary effects and increasing
the lifetime of the dissociated state. The scheme can be, in principle,
coupled to any RD scheme, like over-damped Langevin dynamics, Langevin
dynamics, GFRD \citep{van2005green,ZonTenWolde_PRL05_GFRD} and FPKMC
algorithm \citep{donev2010first}, which could yield additional efficiency
and accuracy or even incorporate long-range hydrodynamic interactions.}

\textcolor{black}{We first implemented the MSM/RD scheme for a simple
ligand diffusion model (Sec. \ref{subsec:3D-diff-pot}), which served
to verify the scheme. It reproduced the expected dynamics and binding/unbinding
rates of the reference simulation. It was also able to generate an
accurate MSM for the internal dynamics with a relatively small amount
of data, which hints that it is feasible to extract the characteristic
dynamics of a computationally feasible amount of MD simulations. Moreover,
we implemented the MSM/RD scheme for the binding of CO to myoglobin
system. After successfully extracting a self-consistent MSM and a
coupling scheme, we found that the equilibrium constant is consistent
with previous experimental and computational results \citep{carver1990analysis,deSancho2015identification}.
We also showed that the MSM/RD estimates are consistent with the underlying
MD simulations – in particular our estimated association rate is consistent
with the association rate estimated from a reference MD simulation
conducted in a large simulation box that was not used to parametrize
the MSM/RD model. This is a significant improvement over Ref. \citep{deSancho2015identification},
where tenfold higher rates were estimated.}

\textcolor{black}{The MSM/RD theory we introduced provides the framework
upon which schemes for more complex systems can be constructed. In
particular, the next steps are to include association of two macromolecules,
which may require to account for rototranslational diffusion, and
the coupling between protein-ligand association and conformational
changes. With the addition of these features, biologically relevant
scenarios can be simulated. For example, if conformational changes
of the protein are rare events and have different ligand association
/ dissociation rates, then the conformational dynamics and the ligand
binding dynamics are nontrivially coupled at high ligand concentrations
– see \citep{PlattnerNoe_NatComm15_TrypsinPlasticity} for the example
of Trypsin and Benzamidine. A biological relevant example is the activation
of the Calcium sensor Synaptotagmin in neuronal synapses \citep{Suedhof_Neuron13_Neurotransmission}.
Here, a locally very high Calcium concentration is created by the
opening of voltage-gated Calcium channels as a response to an electric
signal. Synaptotagmin then binds up to five Calcium ions while going
through different conformations, while the local Calcium concentration
is reduced by diffusion. If Synaptotagmin successfully binds enough
Calcium ions and transitions into an active conformation, it can catalyze
the fission of neuronal vesicles, which transduces the signal to the
postsynaptic side. Such scenarios can be simulated with MSM/RD simulations,
in which the channels, the Synaptotagmin proteins and the ions are
resolved as individual particles, and the binding/dissociation kinetics
and conformational changes of Synaptotagmin is encoded in an MSM.}

\textcolor{black}{MSM/RD could be extended to deal with higher-order
reactions. The most direct approach is to treat interactions of order
2, 3, etc., by different MSMs which are then coupled in a regular
MSM/RD framework. The question then is how the higher-order MSMs are
obtained. The brute-force approach would be to simulate the dynamics
between three or more molecules with MD – e.g. with the help of enhanced
sampling methods – and to extract corresponding higher-order MSMs.
A cheaper, but approximate approach would be to ignore coupling between
different states and assume that multiple ligands can bind and transition
between binding sites independently, perhaps except for multiple occupation
of the same binding site. Based on such an assumption, higher-order
MSMs could be constructed by tensor products of MSMs with one protein
and one ligand. In practice, conducting }\textcolor{black}{\emph{some}}\textcolor{black}{{}
but not all higher-order simulations and combining them to a generative
model via machine learning methods may present a feasible pathway.}

\textcolor{black}{Finally, when considering protein interactions at
high concentrations, the diffusion dynamics and long-range interactions
of proteins are expected to be more complicated and involve hydrodynamic
effects and anomalous diffusion. To include such effects, appropriate
dynamical schemes should be included in the RD part.}

\textcolor{black}{In future developments, we will extend the MSM/RD
scheme to address these issues; however, it should be acknowledged
that some of these extensions come with their own set of challenges
that are not trivial to address.}

\section*{Acknowledgments}

We gratefully acknowledge support by the Deutsche Forschungsgemeinschaft
(grants SFB1114, projects C03 and A04), the Einstein Foundation Berlin
(ECMath grant CH17) and the European research council (ERC starting
grant 307494 \textquotedbl{}pcCell\textquotedbl{}). David De Sancho
was supported by grants CTQ2015-65320- R and RYC-2016- 19590 from
the Spanish Ministry of Economy, Industry and Competitiveness (MINECO).
We also thank Tim Hempel and Nuria Plattner for helpful discussions
and software tutorials.

\section*{Appendix: MSM/RD scheme for Sec. \ref{sec:MSM/RD-implementation}\label{sec:Appendix:MSM/RDscheme}}

\textcolor{black}{Based on the estimated quantities defined in the
Sec. \ref{sec:MSM/RD-implementation}, we introduce an implementation
of the MSM/RD algorithm from Sec. \ref{subsec:MSM/RD-coupled}.}

\texttt{\textcolor{blue}{\noindent}}\texttt{\textcolor{black}{Input: Initial
mode (RD or MSM), initial condition (coordinates $\mathbf{c}_{0}$
or state $s_{0}$, respectively) and $t=0$:}}

\texttt{\textcolor{black}{While $t\leq t_{\mathrm{final}}:$}}
\begin{enumerate}
\item \texttt{\textcolor{black}{If in RD mode:}}
\begin{enumerate}
\item \texttt{\textcolor{black}{Propagate $\mathbf{c}_{t}\rightarrow\mathbf{c}_{t+\tau_{\mathrm{RD}}}$
by diffusion }}
\item \texttt{\textcolor{black}{Update time $t\mathrel{{+}{=}}\tau_{\mathrm{RD}}$}}
\item \texttt{\textcolor{black}{If $r_{AB}(\mathbf{c}_{t})<R$ (enter MSM
domain):}}
\begin{itemize}
\item \texttt{\textcolor{black}{Select trajectory from }}\textcolor{black}{$L_{\mathrm{entry}}=\{\mathbf{c}_{\mathrm{entry}},\mathbf{x}_{\mathrm{end}},\Delta\}$}\texttt{\textcolor{black}{{}
with $\mathbf{c}_{\mathrm{entry}}$ closest to $\mathbf{c}_{t}$}}
\item \texttt{\textcolor{black}{If $\mathbf{x}_{\mathrm{end}}$ is a state:}}~\\
\texttt{\textcolor{black}{Map to state $s_{t+\Delta}=\mathbf{x}_{\mathrm{end}}$}}~\\
\texttt{\textcolor{black}{Update time $t\mathrel{{+}{=}}\Delta$}}~\\
\texttt{\textcolor{black}{Switch to MSM mode}}
\item \texttt{\textcolor{black}{If $\mathbf{x}_{\mathrm{end}}$ are coordinates:}}~\\
\texttt{\textcolor{black}{Map to coordinates $\mathbf{c}_{t+\Delta}=\mathbf{x}_{\mathrm{end}}$}}~\\
\texttt{\textcolor{black}{Update time $t\mathrel{{+}{=}}\Delta$}}
\end{itemize}
\end{enumerate}
\item \texttt{\textcolor{black}{Else (MSM mode):}}
\begin{enumerate}
\item \texttt{\textcolor{black}{If $s_{t}\neq s_{t-\tau_{MSM}}$or previous
mode $\neq$ MSM mode:}}
\begin{itemize}
\item \texttt{\textcolor{black}{Sample exit event with }}\textcolor{black}{$P_{\mathrm{exit},s}$}\texttt{\textcolor{black}{{} }}
\item \texttt{\textcolor{black}{If exit event:}}~\\
\texttt{\textcolor{black}{Uniformly select trajectory from }}\textcolor{black}{$L_{\mathrm{exit},s}=\{\mathbf{c}_{\mathrm{exit}},\Delta\}$}\texttt{\textcolor{black}{}}~\\
\texttt{\textcolor{black}{Map to coordinates $\mathbf{\mathbf{c}_{\mathrm{t+\Delta}}=c}_{\mathrm{exit}}$}}~\\
\texttt{\textcolor{black}{Update time $t\mathrel{{+}{=}}\Delta$}}~\\
\texttt{\textcolor{black}{Switch to RD mode and break current loop
iteration}}
\end{itemize}
\item \texttt{\textcolor{black}{Propagate $s_{t}\rightarrow s_{t+\tau_{\mathrm{MSM}}}$
using}}\textcolor{black}{{} $\mathbf{T}(\tau_{\mathrm{MSM}})$}
\item \texttt{\textcolor{black}{Update time}}\textcolor{black}{{} $t\mathrel{{+}{=}}\tau_{\mathrm{MSM}}$.}
\end{enumerate}
\end{enumerate}
\textcolor{black}{The diffusion in the RD domain is done using a Euler-Maruyama
discretization of Eq. (\ref{eq:COM_diffusion}) \citep{higham2001algorithmic}.
Note the diffusion step can be simulated more efficiently with event-based
algorithms, like FPKMC or eGFRD \citep{donev2010first,van2005green,takahashi2010spatio,vijaykumar2015combining}
for systems with low particle concentrations. In order to optimize
the efficiency of the algorithm, the entry points of entry trajectories
are classified into equal area bins on the sphere. This allows the
algorithm to find the closest trajectory to a given entry point more
efficiently. The partition of the sphere was done following \citep{leopardi2006partition}.}

\bibliographystyle{unsrtnat}
\bibliography{literature,all,own}

\end{document}